\documentclass[usenatbib]{mn2e}
\input epsf
\usepackage{epsfig}
\usepackage{amsmath}

\def\plotone#1{\centering \leavevmode
\epsfxsize=\columnwidth \epsfbox{#1}}
\def\plottwo#1#2{\centering \leavevmode
\epsfxsize=.99\columnwidth \epsfbox{#1} \hfil
\epsfxsize=.99\columnwidth \epsfbox{#2}}

\def\spose#1{\hbox to 0pt{#1\hss}}

\def\approxlt{{\mathrel{\spose{\lower 3pt\hbox{$\sim$}}
        \raise 2.0pt\hbox{$<$}}}}
\def\approxgt{\mathrel{\spose{\lower 3pt\hbox{$\sim$}}
        \raise 2.0pt\hbox{$>$}}}

\newcommand{\drv}[2]{\frac{\partial #1}{\partial #2}   }

 \title[Effect of turbulent diffusion on iron abundance profiles]{Effect of turbulent diffusion on iron abundance profiles}

\author[P.~Rebusco et al.]{P.~Rebusco $^{1}$, E.~Churazov$^{1,2}$, H.~B\"ohringer$^{3}$, W.~Forman$^{4}$ \\
$^1$ Max-Planck-Institut f\"ur Astrophysik, Karl-Schwarzschild-Strasse 1, 85741
Garching, Germany\\
$^2$ Space Research Institute (IKI), Profsoyuznaya 84/32, Moscow 117810, 
Russia\\
$^3$ MPI f\"{u}r Extraterrestrische Physik, P.O. Box 1603, 85740
Garching, Germany\\
$^4$ Harvard-Smithsonian Center for Astrophysics, 60 Garden St.,
Cambridge, MA 02138, USA
}

\pagerange{\pageref{firstpage}--\pageref{lastpage}}
\pubyear{2005}
\usepackage{subfigure}
\begin{document}

\maketitle
\label{firstpage}

\begin{abstract}
We compare the observed peaked iron abundance profiles for a small
sample of groups and clusters with the predictions of a simple model
involving the metal ejection from the brightest galaxy and the subsequent
diffusion of metals by stochastic gas motions. Extending the analysis
of Rebusco et al. (2005) we found that for 5 out of 8 objects in the
sample an effective diffusion coefficient of the order of $10^{29}$
cm$^{2}$~s$^{-1}$ is needed.  For AWM4, Centaurus and AWM7 the results
are different suggesting substantial intermittence in the process of
metal spreading across the cluster. There is no obvious dependence of
the diffusion coefficient on the mass of the system.

We also estimated the characteristic velocities and the spatial scales
of the gas motions needed to balance the cooling losses by the
dissipation of the same gas motions. A comparison of the derived
spatial scales and the sizes of observed radio bubbles inflated in the
ICM by a central active galactic nucleus (AGN) suggests that the
AGN/ICM interaction makes an important (if not a dominant)
contribution to the gas motions in the cluster cores.
\end{abstract}


\section{Introduction}
 X-ray spectroscopy is widely used to determine the metallicity of
the hot gas in galaxy clusters and groups. Using the high energy
resolution of ASCA, \textit{BeppoSAX}, \textit{Chandra} and
\textit{XMM-Newton} the radial profiles of Fe, S, Si, Ca and other
elements (e.g Buote et al. 2003b, Sanderson et al. 2003, Sun et
al. 2003, Matsushita, Finoguenov \& B\"ohringer 2002, Gastaldello\&
Molendi 2002, O'Sullivan et al. 2005, Fukazawa 1994, Tamura et
al. 2001) have been derived for
a sample of objects. Outside the cluster core regions the metallicity
of the intracluster medium (ICM) is on average one-third of the solar
one and it does not seem to evolve up to a redshift of about $1$ (e.g
Mushotzky \& Loewenstein 1997, Tozzi et al. 2003). This suggests an
early enrichment by Type II supernovae (SNII; e.g. Finoguenov et
al. 2002). The metallicities in the core regions demonstrate much more
diversity. For the so-called 'non-cooling flow' clusters the
metallicity and the surface brightness do not vary strongly across the
core region. For another group of so-called 'cooling flow' clusters, the
metallicity and the surface brightness are both strongly peaked at the
center. The clusters from the latter group always have a very bright
galaxy (BG) dwelling in their centers, which makes these galaxies a
prime candidate for producing the peaked abundance profiles.  These
peaked profiles were likely formed well after the cluster/group was
assembled (e.g. B\"ohringer et al. 2004; De Grandi et al. 2004).  The
relative abundances of different elements indicate that Type Ia
supernovae (SNIa) explosions and stellar winds within the BG have
played a major role in the formation of the central abundance excesses
(e.g. Renzini et al. 1993, Finoguenov et al. 2002).  If the metals in the
ICM are due to the galaxy stars then one would expect the abundance
profiles to follow the BG light profile. However the observed metal
profiles are much less steep than the light profiles, suggesting
that there must be a mixing of the injected metals, a process which
may help to diffuse them to larger radii (e.g.Churazov et al. 2003,
Rebusco et al. 2005, Chandran 2005). Stochastic gas motions could provide such a
mechanism of spreading metals through the ICM, provided that the
characteristic velocities and the spatial scales are of the right
order. Furthermore, the dissipation of the kinetic energy of the same
gas motions could be an important source of energy for the rapidly cooling
gas in the ``cooling flow'' clusters.

In Rebusco et al. 2005 we estimated the parameters (velocity and
length scale) of stochastic gas motions in the core of Perseus cluster
(A426) required to spread the metals ejected from the central galaxy.  We
found that a diffusion coefficient of the order of $2\times10^{29}$
cm$^2$~s$^{-1}$ is needed to explain the observed abundance profile
and that for turbulent velocities of about $300$ km~s$^{-1}$ and
eddies sizes of about $20$ kpc the dissipation of turbulent motions
compensates for the gas cooling losses. We now extend the same model
to a small sample of clusters and groups. The sources we analyzed have
mean temperatures which vary by a factor of $\sim 6$.


The structure of the paper is the following.  In section
\ref{sec_sample} we describe the sample of clusters and groups; in
section \ref{sec_model} the structure and the ingredients of the
stochastic diffusion model are described. The results are discussed in
section \ref{sec_results}. The last section summarizes our findings.\\
We adopt a Hubble constant of $H_0=70$ km~s$^{-1}$~Mpc$^{-1}$,
$\Omega_M=0.3$ and $\Omega_\Lambda=0.7$.

\section{The Sample}
\label{sec_sample}
We have selected a small sample of nearby ``cooling flow'' clusters
and groups ($z < 0.07$) having a sufficiently detailed information on
the abundance distributions in the core regions. To test for any obvious
trend with the mass of the cluster, we selected the objects having the
mean temperature ranging from $1$ to $6$ keV  (see Table
\ref{tab_sample}).


\begin{table}
\begin{center}
\begin{tabular}{c |c c }	
\hline
\hline
Name& $kT$    & $z$  \\
\hline
NGC 5044     &  $1$   &  $0.0082$          \\
NGC 1550     & $1.37$ & $0.0124$           \\	
M87          & $2.2$  & $0.0044$          \\
AWM4         & $2.3$  & $0.0319$         \\
Centaurus    & $3.5$  & $0.0107$\\ 
AWM7         & $3.7$  & $0.0176$             \\
A1795        & $6$    & $0.0639$      \\
Perseus      & $6.3$  & $0.0179$\\
\hline
\end{tabular}
\caption{The sample: (1) name of the cluster , (2) mean ICM
temperature in keV, (3) redshift.}
\label{tab_sample}
\end{center}
\end{table}
 

\subsection{NGC 5044 Group}
The NGC 5044 group of galaxies is relatively cool and loose. It is
formed by a central luminous giant elliptical surrounded by about 160
galaxies, $\sim80\%$ of which  are dwarfs (Ferguson \& Sandage 1990).
There is evidence for multiple temperature gas components, that
coexist within $\sim 30$ kpc. It is likely that the central heating is
due to the presence of a central AGN (e.g. Buote et al. 2003). The
X-ray emission is quite symmetric, indicating that there have not been
recent strong mergers (David et al. 1994).

\subsection{NGC 1550 Group}
This luminous group is more relaxed than other bright low temperature
ones (e.g. NGC 5044). The BG is a lenticular galaxy and it is almost
isolated in the optical band (Garcia 1993). \\ Sun et al. (2003)
suggest that NGC 1550 entropy profile shows signs of
nongravitational heating (e.g. a recent outburst).  Moreover they
point out the large role of the central galaxy in affecting the
surrounding gas temperature.\\ If this is the case, then the same
nongravitational processes may help to distribute the metals within
the group core.

\subsection{M87 Galaxy} 
M87 is the dominant central galaxy in the Virgo Cluster. Because M87
is very close and bright it was used to study the role of SNIa and
SNII in enriching the ICM (e.g. Gastaldello \& Molendi 2002,
Matsushita et al. 2003, Finoguenov et al. 2002).

There is clear dynamical evidence of the presence of a supermassive
black hole in its core (e.g. Macchetto et al. 1997), with a one-sided
jet (e.g. Schreier et al.1982, Owen et al. 1989, Sparks et al. 1996).
  At
least two gas components with different temperatures are observed: the
hotter component is almost symmetric around the central galaxy
(Ma\\tsushita et al. 2002), while the cooler component forms extended
structures correlated with the radio lobes. There are clear signs of
an AGN/ICM interaction in this source both through the shocks and sound
waves (Forman et al. 2005, 2006) and through mechanical entrainment of
the cool gas by the bubbles of relativistic plasma (Churazov et
al. 2001).

\subsection{AWM4 Cluster}

This poor relaxed cluster consists of 28 galaxies centered at the
dominant elliptical NGC 6051 (Koranyi \& Geller 2002). The cluster's
stellar component is comparable with that of M87 and NGC 5044, but the
iron abundance profile is flatter than in these sources and there is
no evidence for multiphase gas (O'Sullivan et al. 2005). The density
is not strongly peaked at the center and no strong temperature drop
is seen. While it is not a prominent cooling flow cluster, it does
contain a single cD galaxy and we include this cluster to see what will
be the outcome of the analysis applied to this object. 

The central galaxy hosts a powerful AGN (4C +24.36) which could be
driving the gas motions in this cluster.\\ O'Sullivan et al. (2005) assert
  that in AWM4 there might be global motions of the galaxies in 
 the plane of the sky: these galaxy motions could contribute in
 spreading the metals.

\subsection{Centaurus Cluster}
Centaurus is the third nearest bright cluster (after Perseus and
Virgo). The asymmetric X-ray structure around the central cD galaxy
(NGC 4696) gives evidence for dynamical activity (e.g. Allen \& Fabian
1994).  This cluster is very interesting because it houses the most
prominent abundance peak known (e.g. Fabian et al. 2005, Sanders \&
Fabian 2002, Ikebe et al. 1999).  The analysis of the cool H$\alpha$
filaments in its core indicates previous episodes of radio activity
(Crawford et al. 2005).  Fabian et al. 2005 and Graham et al. 2006
have recently calculated the effective diffusion coefficient in the
picture of stochastic turbulence: we discuss our results in comparison
to these earlier works in \ref{subsec:cen}.

\subsection{AWM7 Cluster}
AWM7 is a poor bright cluster, part of the Perseus-Pisces
super-cluster and centered about the dominant elliptical galaxy NGC
1129 (e.g. Neumann \& B\"ohringer 1995). It is elliptically elongated
in the east-west direction and it follows the general orientation of
the Perseus-Pisces chain of galaxies (Neumann \& B\"ohringer
1995). The $30$ kpc shift of the X-ray peak from the optical peak
indicates that AWM7 has a cD cluster in early stage of evolution (Furusho et
al. 2003). The determination of the effective radius of NGC 1129 is
quite controversial (see Table $2$).  This uncertainty does not affect
our estimates, however.
     
\subsection{A1795 Cluster}
This compact and rich cluster is believed to being dyna\\
mically relaxed (e.g.Tamura et al.2001). On the other hand Ettori et al.(2002)
suggest that the core has an unrelaxed nature, consistent with the
detected motion of the central cD galaxy MCG +5-33-5 (Oegerle \& Hill
1994, Fabian et al. 1994, Markevitch et al. 1998). There is a strong cooling flow in this cluster with
a substantial temperature drop and a strong density peak in the core
(e.g. Briel \& Henry 1996, Allen et al. 2001).
     
\subsection{Perseus Cluster}
Perseus (A 426) is the brightest nearby X-ray cluster and it is one of
the best-studied cases of cool core clusters, together with M87 and
Centaurus.  Its central elliptical galaxy (NGC 1275) dominates in the
optical light up to a distance of $\sim 100$ kpc.  It hosts a
moderately powerful radio source, 3C 84 (Pedlar et al. 1990).  In the
core region there is a complex substructure in the temperature and
surface brightness di\\
stributions including depressions in surface
brightness due to rising bubbles of relativistic plasma (Churazov et
al., 2000, Fabian et al. 2000) and quasi-spherical ripples (Fabian et
al. 2003a, 2006). Optical H$\alpha$ filaments, whose origin is not
well understood, seem to be drawn up by the rising bubbles (Fabian et
al. 2003b, Hatch et al. 2005).

\section{The model}
The model is essentially the same used in Rebusco et
al. (2005). For completeness we reproduce its main features
below.

\label{sec_model}
\subsection{Diffusion of metals due to stochastic gas motions}
We assume a static gas distribution in the gravitational potential of
the cluster. The gas density and temperature di\\
stributions are known
and assumed to be not evolving with time. We then suppose that the gas
is involved in stochastic motions which do not affect the density
and temperature distributions on  time scales much longer than the
time scales associated with the gas motions, but only spread the
metals through the ICM. Due to such motions, the metals injected in
the center of the cluster will be spread out off the center. We treat
this process in the diffusion approximation:
\begin{equation}
\drv{n a}{t}=\nabla \cdot (D n \nabla { a} )+ S,
\label{eq:dif}
\end{equation}
where $n=n(r)$ is the gas density, $a=a(r,t)$ is the iron abundance
and $S=S(r,t)$ is the source term due to the iron injection from the
BG. $D$ is the diffusion coefficient, of the order of
$\sim\frac{vl}{3}$, with $v$ being the characteristic velocity of the
stochastic gas motions and $l$  their characteristic length scale.
Once the diffusion coefficient is
specified, eq.\ref{eq:dif} can be integrated.  In what follows we
considered constant diffusion coefficients: the effect of having a
diffusion coefficient as a function of radius is discussed in Rebusco
et al. (2005).

\subsection{Adopted light, gas density and metal abundance profiles}
\label{subsec_profiles}
We list the parameters of the adopted profiles in Table 2: we used the
existing analytical approximations (when available) or made our own
fits to the original data (all the references used are given in the Table).

The light distribution of the central cD galaxies is modeled here by a
simple Hernquist profile (Hernquist 1990): this is an acceptable
approximation for the purpose of this study. For each source we
compared the light distribution of the BG with the light distribution
due to all the other gala\\
xies excluding the BG. The brightest galaxy
dominates up to a distance of $\sim$ 100-150 kpc, hence in the
calculations we assume (unless explicitly stated otherwise) that the
central excess of metals is produced by the central galaxy alone.

The iron abundance profile is approximated with a simple
$\beta$-profile: $a(r)=a(0)\left[1+(r/r_a)^2 \right]^{-b}$, where
$a_\odot$ is the solar abundance (Anders \& Grevesse 1989). The
adopted functional form should work only in the central region, where
the abundance excess is present.  This form also neglects completely
the central abundance "hole'' observed in some sources (e.g. Schmidt
et al. 2002, B\"ohringer et al. 2001). The nature of this abundance
hole remains unexplained and it may be due to a not adequate modeling
of the emission from the very central region rather than due to a real
decrease of the metal abundance.

For the electron density profile we used a single $\beta$ model again:
$n_e(r)=n(0)\left[1+(r/r_c)^2 \right]^{-3/2 \beta}$. The hydrogen
number density is assumed to be related to the electron number density
as $n_{H}=n_{e}/1.2$. In some cases (marked with an asterisk) a double
$\beta$-profile is reported in the original papers: here we use the
parameters of the component which provides the dominant contribution
for the range of radii of interest.  Only in the case of Centaurus the
double profile have been used, because both the components are
important.  For Perseus cluster we adopted the same profiles as in
Rebusco et al. (2005).


\subsection{Iron enrichment}
\label{subsec_source}
Much of the metals in the cluster ICM were produced at early times by
SNII explosions.  They are likely to be evenly distributed through the
bulk of the ICM producing a uniformly enriched gas. The level of this
uniform enrichment is uncertain - on average an iron abundance in the
range 0.2--0.4 of the the solar value is reported. In the subsequent
calculations we subtracted this "abundance basis" $a_b$ in order to
single out only the central abundance excess, which is believed to be
formed later by the metal ejections from the BG alone (Matsushita et
al. 2003, B\"ohringer et al. 2004, De Grandi et al. 2004). The assumed
value of $a_b$ is one of the main uncertainties in our model - for low
values of $a_b$ the central abundance excess becomes more extended and
the mass of  iron in the central excess grows substantially.

Ram-pressure stripping can also contribute to the enrichment of the
ICM over a long period of cluster evolution. According to recent
simulations (Domainko et al. 2006) the cluster centers are enriched
more strongly than the outer parts. It is not clear though how these
results will be affected by the presence of  cool and dense
cores. Below we assume that ram pressure stripping produces broader
abundance peaks than considered here and that the central abundance excess
is solely due to the central galaxy.

We assume that the central abundance excess is due to the type Ia
supernovae and the stellar mass loss by stars in the BG. 
Both channels inject gas enriched with heavy elements, in
particular iron. Following B\"ohringer et al. (2004), the rates of
iron injection by SNIa (eq. 2) and stellar mass loss (eq. 3) are:
\begin{eqnarray}
\left (\frac{dM_{Fe}}{dt} \right )_{SN Ia} & = & SR \times10^{-12}
\left (\frac{L_B}{L_{\odot}^B} \right ) \eta_{Fe} \\ & = & R(t)~
0.105\times 10^{-12} \left ( \frac{L_B}{L_{\odot}^B} \right )
M_{\odot}~yr^{-1},\nonumber \\ \left ( \frac{dM_{Fe}}{dt} \right
)_{*}& = &\gamma_{Fe}\times 2.5 \times 10^{-11} \left (\frac{t}{t_{H}}
\right )^{-1.3} \\ \nonumber & & \times \left (
\frac{L_B}{L_{\odot}^B} \right ) M_{\odot}~yr^{-1},
\label{eq:rate}
\end{eqnarray}
where $SR$ is the present SNIa rate in SNU (1 SNU - rate of supernova
explosions, corresponding to one SN event per century per galaxy with
a blue luminosity of $10^{10}~L_{\odot}^B$), $\eta_{Fe}=0.7 M_{\odot}$
is the iron yield per SNIa, $\gamma_{Fe}=2.8\times10^{-3}$ is the mean
iron mass fraction in the stellar winds of an evolved stellar
population, $t_H$ is the Hubble time. The expression for the stellar
mass loss was adopted from Ciotti et al. (1991), assuming a galactic
age of $10$ Gyr. The stellar mass loss contribution to the hydrogen
content of the ICM can be neglected, as its effect is important only
in the central $\sim 10-20$ kpc (see Fig. $1$).  The time dependent
factor $R(t)= (t/t_{H})^{(-k)}$ takes into account an increased SNIa
rate in the past (Renzini et al. 1993), with the index $k$ ranging
from $1.1$ up to $2$.  We assume a fiducial value for the present day
SNIa rate of $0.15$ SNU (Cappellaro, Evans, \& Turatto 1999). Hence
the total iron injection rate within a given radius $r$ can be written
as
\begin{eqnarray}
s(<r,t)=\left ( \frac{dM_{Fe}}{dt} \right )_{*}+\left (
\frac{dM_{Fe}}{dt} \right )_{SN} \propto \left (
\frac{L_B(<r)}{L_{\odot}^B} \right )
\label{eq:source}
\end{eqnarray}

The procedure of evaluating the diffusion coefficient in equation (1)
was as follows. First the value of $a_b$ was spe\\
cified and subtracted
from the observed abundance profile. The iron mass in the remaining
abundance excess was then calculated and compared with the total
amount of iron produced by the central galaxy during its evolution
from some initial time to the present (by integrating equation 4 over
time). This provides the constraints on the parameters of the
enrichment model ($SR$, $k$, $t_{age}$) (see Rebusco et al. 2005 for
an extended discussion). For each source we explored diffe\\rent models,
with $a_b\in\left[0.2,0.4\right] $, $k\in\left[1.1,2\right] $ $SR \in
[0.07,0.34] $ and $t_{age}\in [5,10]$. In particular, we fixed
$a_b=0.2, 0.3$ and $0.4$; then for each abundance basis we found a
sample of enrichment parameters: e.g. by setting $k=1.1,2$ and
$SR=0.15,0.3$, and calculating the $t_{age}$ necessary to get the
right amount of iron excess, then by fixing $SR$ and $t_{age}$ and
calculating $k$, finally by calculating $SR$ from four combinations of
$k$ and $t_{age}$.  A representative set of these models is listed in
Table $4$.

Equation (1) was then integrated over time from $t_H-t_{age}$ till now
starting from a zero initial abundance. The resulting profiles were
visually compared with the observed abundance peaks and the diffusion
coefficient was adjusted so that the predicted and observed profiles
would agree reasonably well (see Fig.1). In this figure for each object in
the sample the dashed line shows the expected iron abundance due to
the ejection of metals from the galaxy if no mixing is present. In
most cases such profiles are far too steep than the observed abundance
excesses (shown with a solid line). The dotted line in Fig.1 shows the
expected profiles calculated for the same parameters of iron injection
model, but with the additional effect of diffusion. The actual
values of the diffusion coefficient used in each case are listed in
Table 3. Clearly allowing for metal mixing leads to a reasonable
agreement between the observed and the predicted abundance profiles.

\subsection{Cooling and heating}
\label{subsec_vl}

As shown above the observed and predicted abundance profiles can agree
reasonably well if the metals are allowed to diffuse through the
ICM. Assuming that the diffusion coefficient derived above is due to
stochastic gas motions, one can cast it in a form $D\sim C_1 vl$, where
$C_1$ is a dimensionless constant of the order of unity, $v$ is the
characteristic velocity of the gas motions and $l$  their
characteristic length scale. Thus the above analysis provides an
estimate of the product of $v$ and $l$ for each object.

\subsubsection{Dissipation of turbulent motions}
Here we further assume that the dissipation of turbulent motions is the
dominant source of energy for the rapidly cooling gas in the cluster
cores. As in Rebusco et al. 2005 we suppose that the dissipation rate
is simply equal to the gas cooling rate
$\Gamma_{diss}\sim\Gamma_{cool}$.  The gas heating rate due to the
dissipation of the kinetic energy can be written as $\Gamma_{diss}\sim
C_2\rho v^3/l$, where $\rho$ is the gas density and $C_2$ is a
dimensionless constant, while the cooling rate
$\Gamma_{cool}=n^2\Lambda(T)$ can be estimated from the observed gas
temperature and density at any given radius.

This requirement of the cooling and heating losses provides a second
constraint on the combination of $v$ and $l$. Thus fixing the diffusion
coefficient and the heating rate one can easily estimate $v$ and $l$, that we intend as the 
characteristic sizes and velocities of quasi-continua intermittent vortexes. 
 
Below we use the expressions for $D$ and $\Gamma_{diss}$ from Dennis \&
Chandran (2005) (see references therein):
\begin{eqnarray}
D&=&C_1 v l  \sim 0.11 v l \\
\Gamma_{diss}&=& C_2 \rho v^3/l \sim  0.4 \rho v^3/l. 
\label{eq:dd}
\end{eqnarray}
To evaluate the heating rate for each object from the sample we use the
gas parameters near $r_0=r_{cool}/2$ (where $r_{cool}$ is the radius
at which the cooling time is comparable with the Hubble time). We then
used $D$ from Table 3 and set $\Gamma_{diss}=n^2\Lambda(T)$, evaluated
at $r_0$.

\onecolumn
\begin{figure}
\centering
    \subfigure[NGC 5044]{
     \label{fig:ab5044}
      \includegraphics[width=5.4cm]{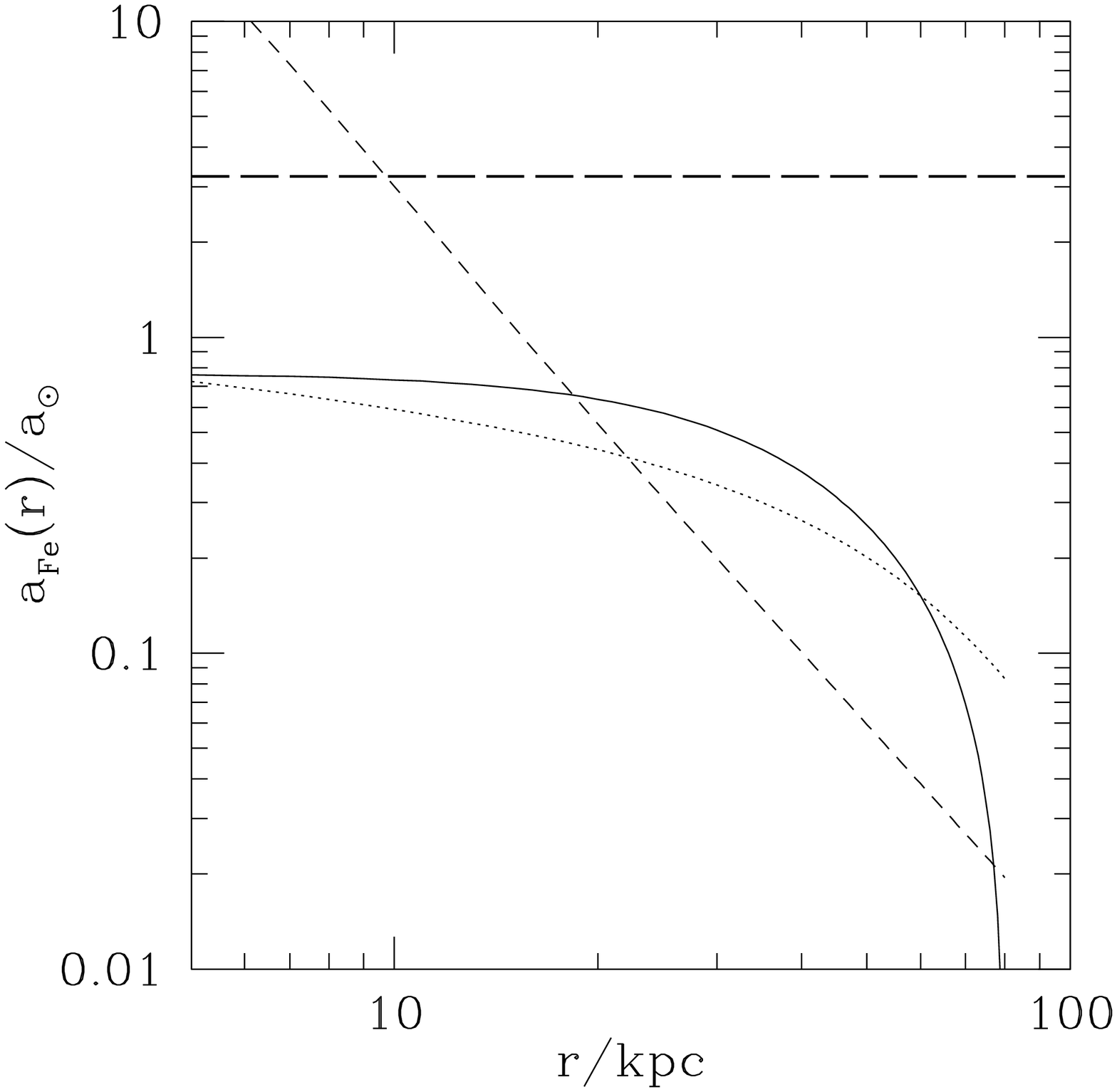}}
   \subfigure[NGC 1550]{
   \label{fig:ab1550}
      \includegraphics[width=5.4cm]{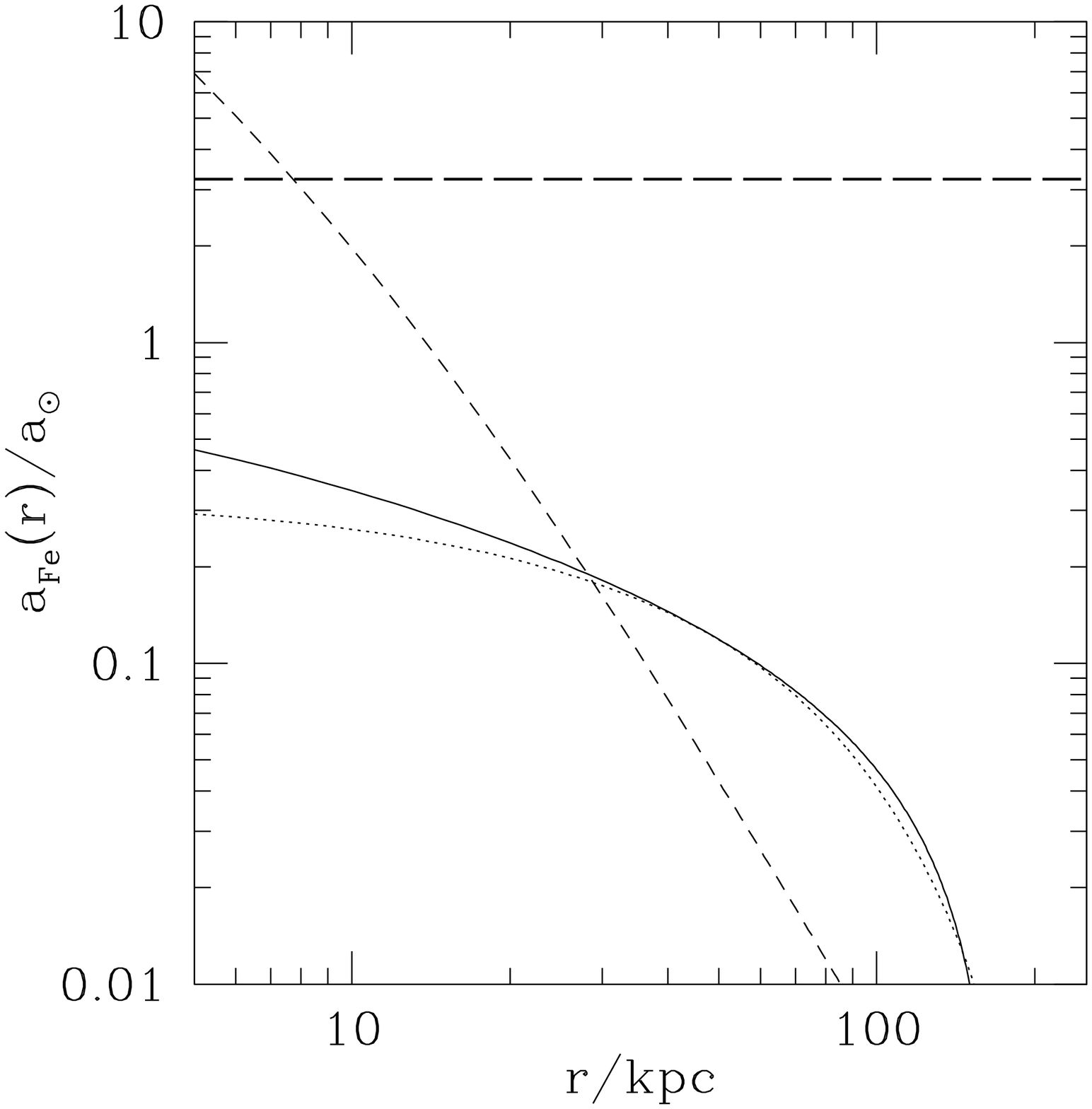}}
    \subfigure[M87]{
    \label{fig:abM87}
     \includegraphics[width=5.4cm]{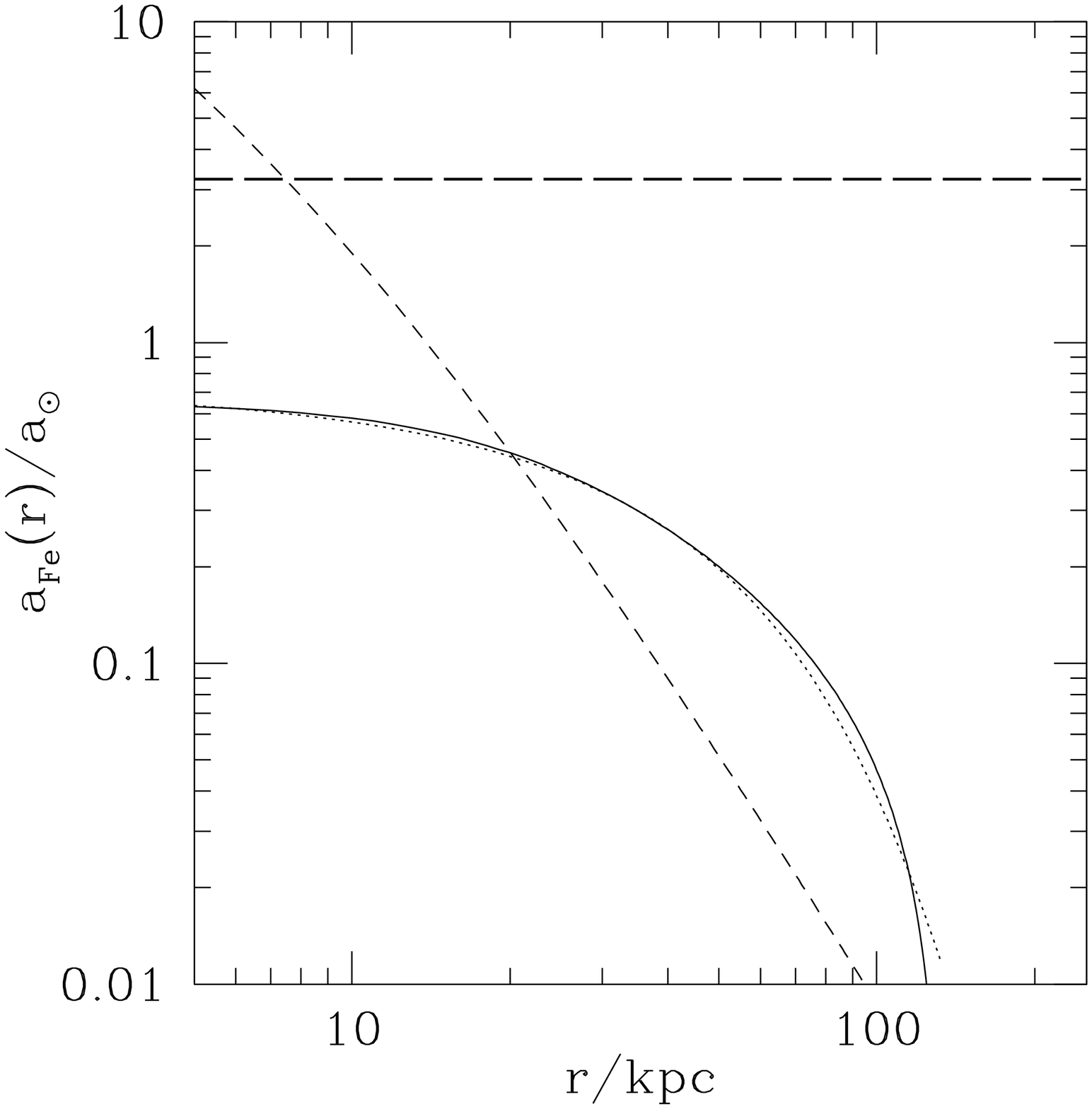}}\\
    \subfigure[AWM4]{
    \label{fig:abAWM4}
     \includegraphics[width=5.4cm]{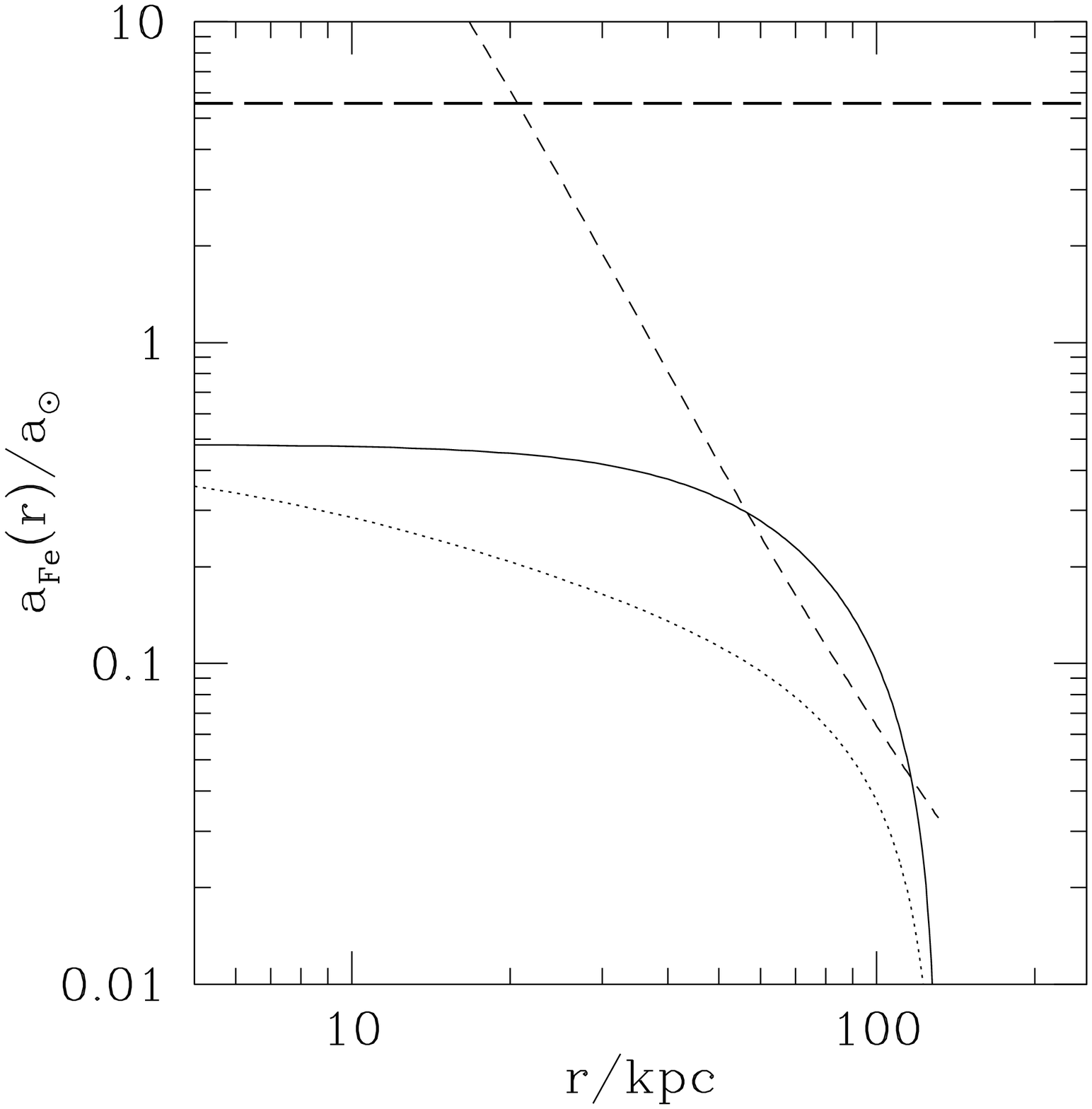}}
    \subfigure[Centaurus]{
    \label{fig:abcen}
     \includegraphics[width=5.4cm]{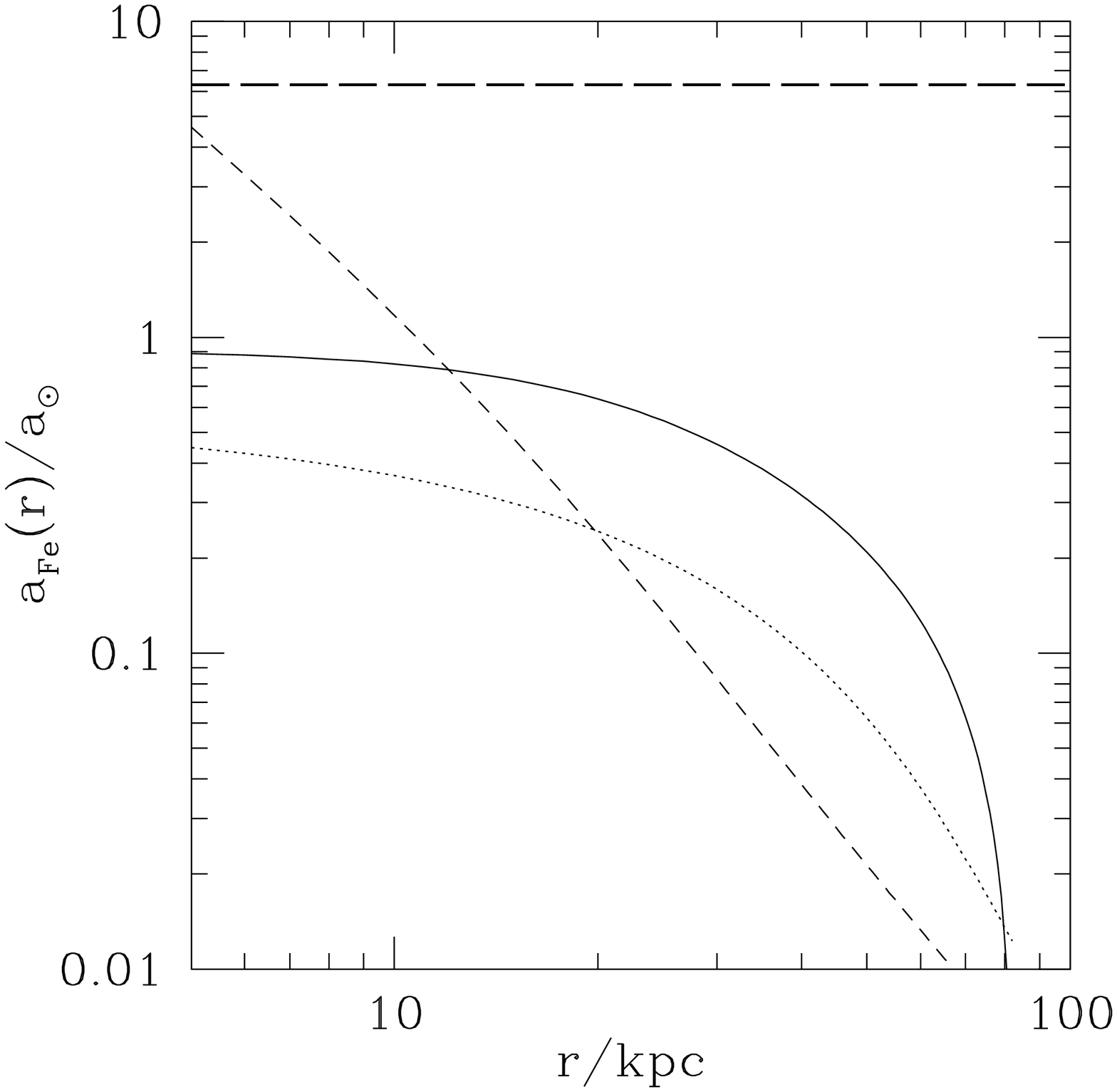}}
    \subfigure[AWM7]{
    \label{fig:abAWM7}
     \includegraphics[width=5.4cm]{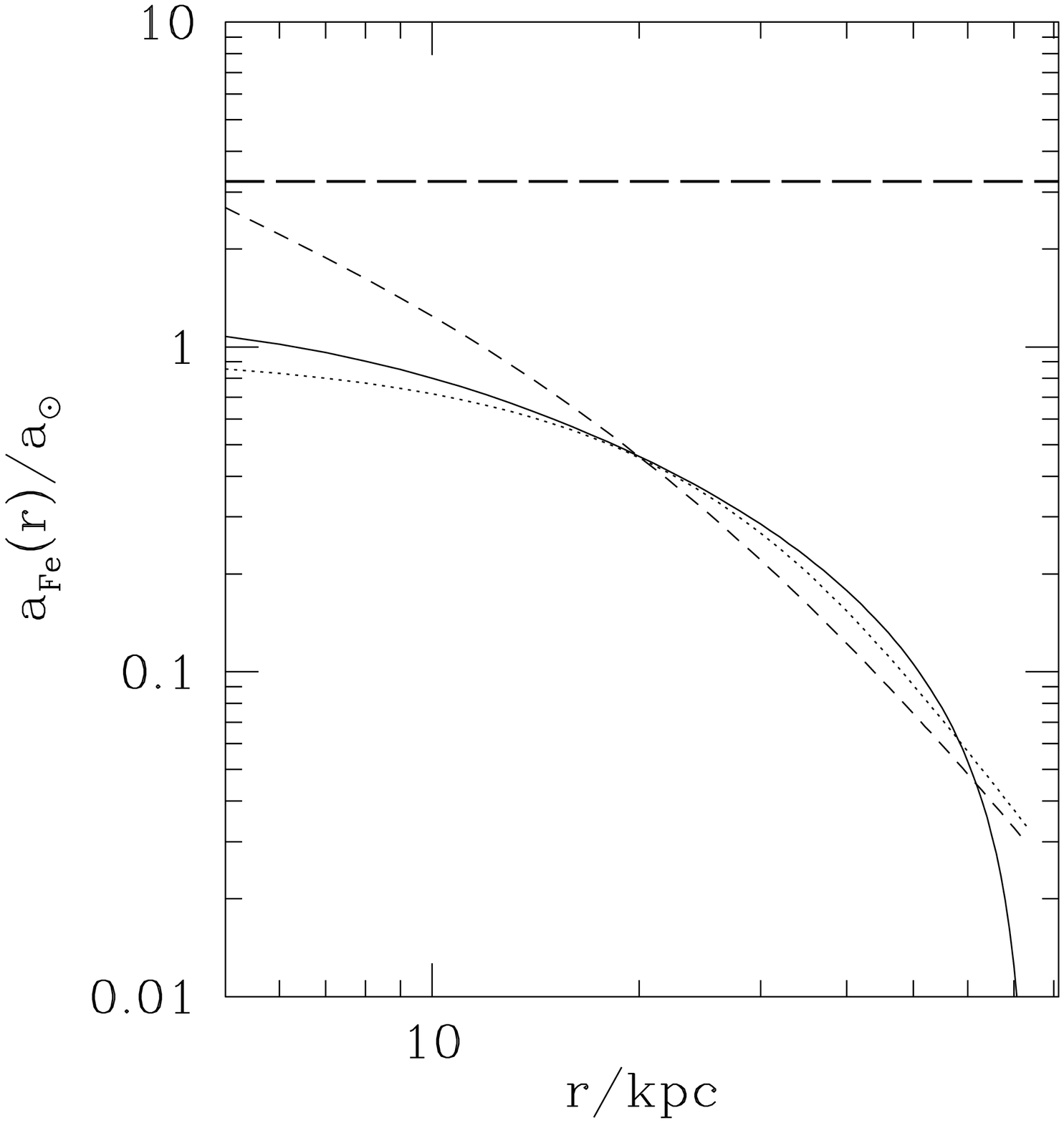}}\\
    \subfigure[A1795]{
    \label{fig:ab1795}
     \includegraphics[width=5.4cm]{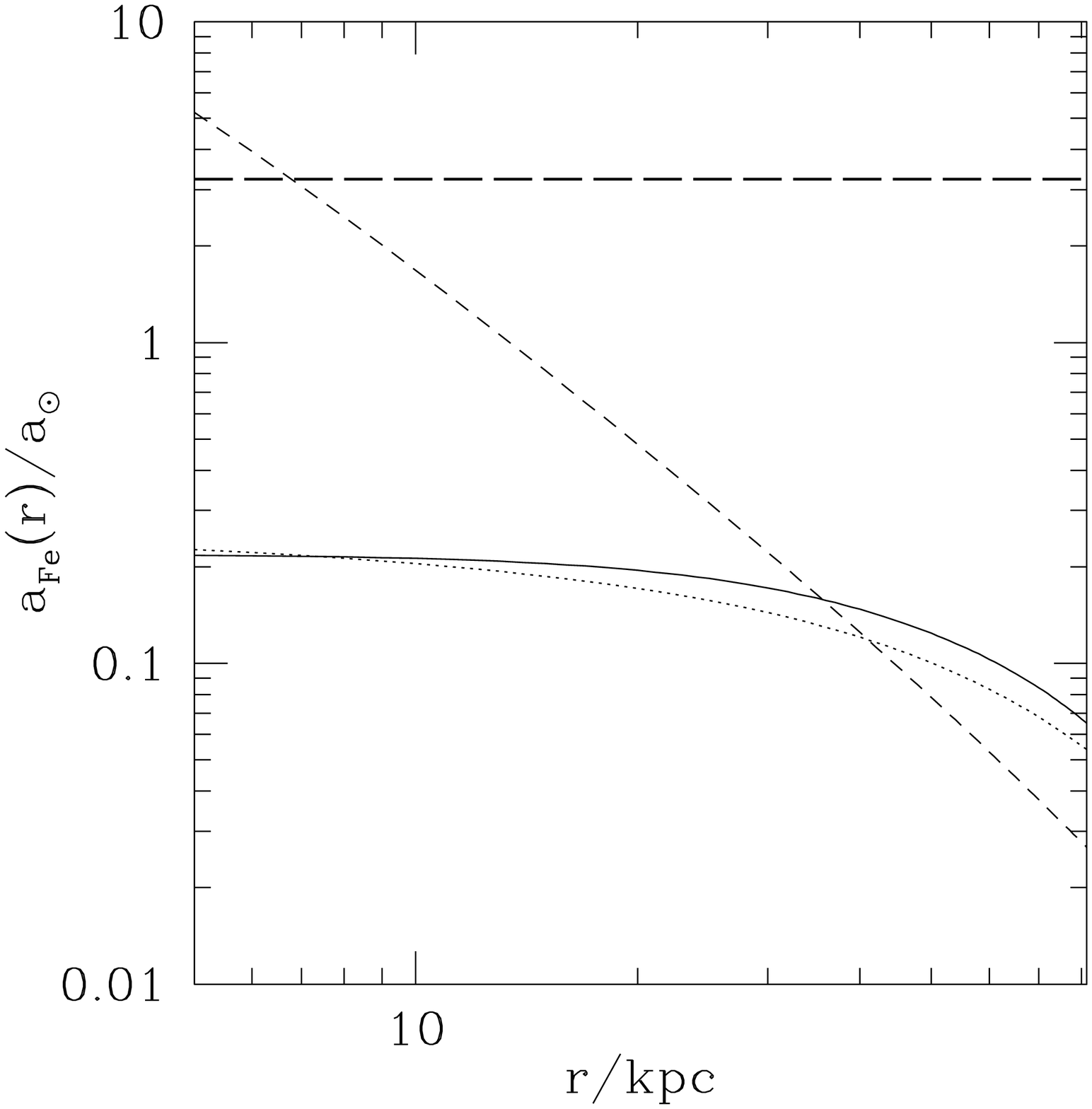}}
    \subfigure[Perseus]{
    \label{fig:abperseus}
     \includegraphics[width=5.4cm]{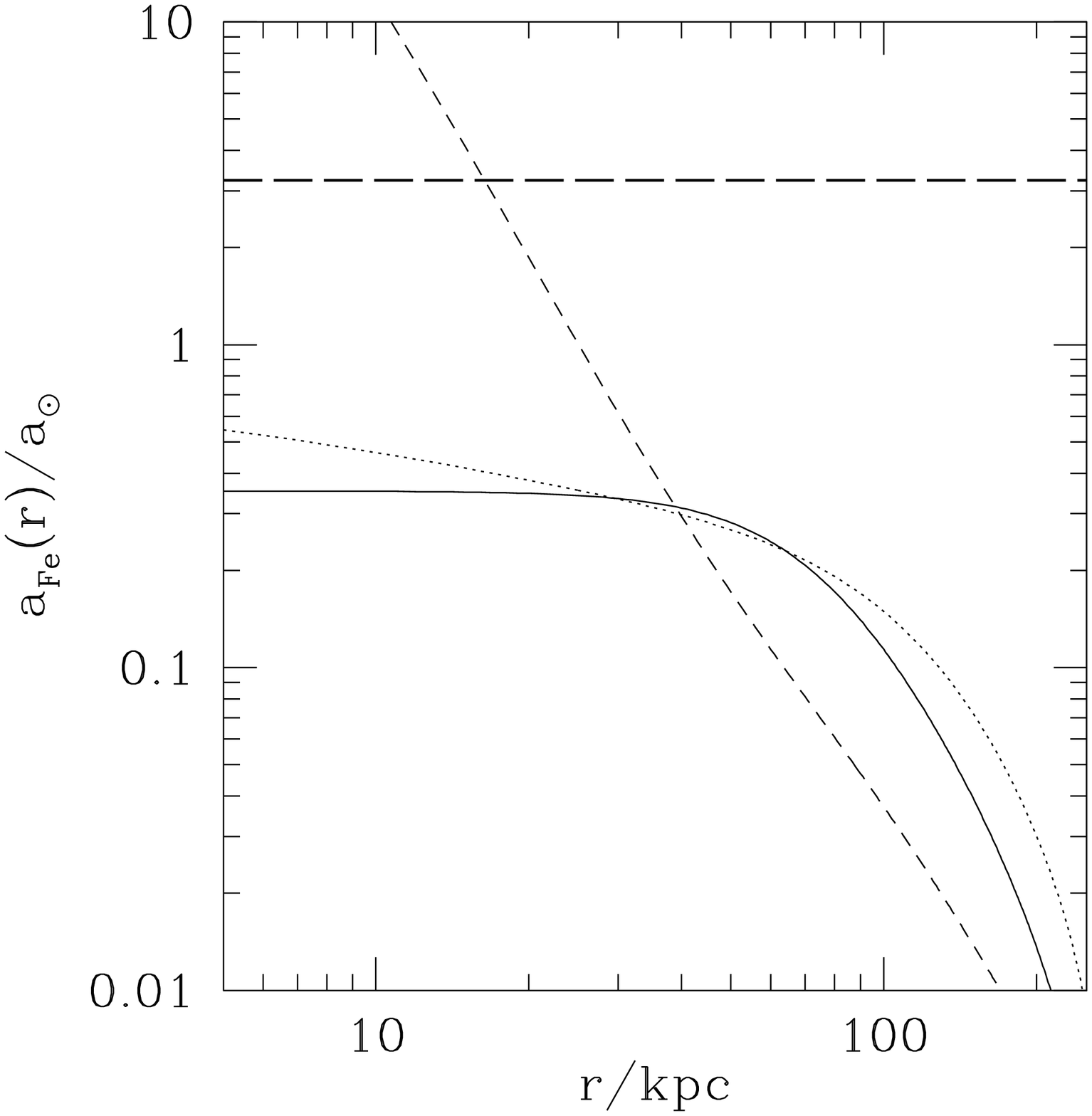}}
\caption{Comparison of the observed and the expected iron abundance
profiles for each source: the set of enrichment parameters used are
listed in Table $4$.  The solid line shows the abundance profile
adopted in this paper, from which a constant value of $a_b$ is
subtracted.  We assume that this central abundance excess is mainly
due to the metal ejection of the central galaxy. For comparison we
show the expected iron abundance (short dashed line) due to the
ejection of metals from the galaxy. The expected profile was
calculated assuming that the ejected metal distribution follows the
optical light.  In all the objects the expected abundance profile is
much more peaked than the observed profile (due to the contribution of
the central galaxy) suggesting that some mechanism is needed to spread
the metals.  The dotted line shows the profile derived with the same
parameters of iron injection, but with the additional effect of
diffusion.  The long dashed line corresponds to the maximum abundance
(it can be obtained from eq. $4$), beyond which our approximation of
neglecting the gas injection is not valid anymore.  }
\label{fig:1multifig}
\end{figure}

\twocolumn
\onecolumn
\begin{figure}
\centering
    \subfigure[NGC 5044]{
     \label{fig:lv5044}
      \includegraphics[scale=0.27]{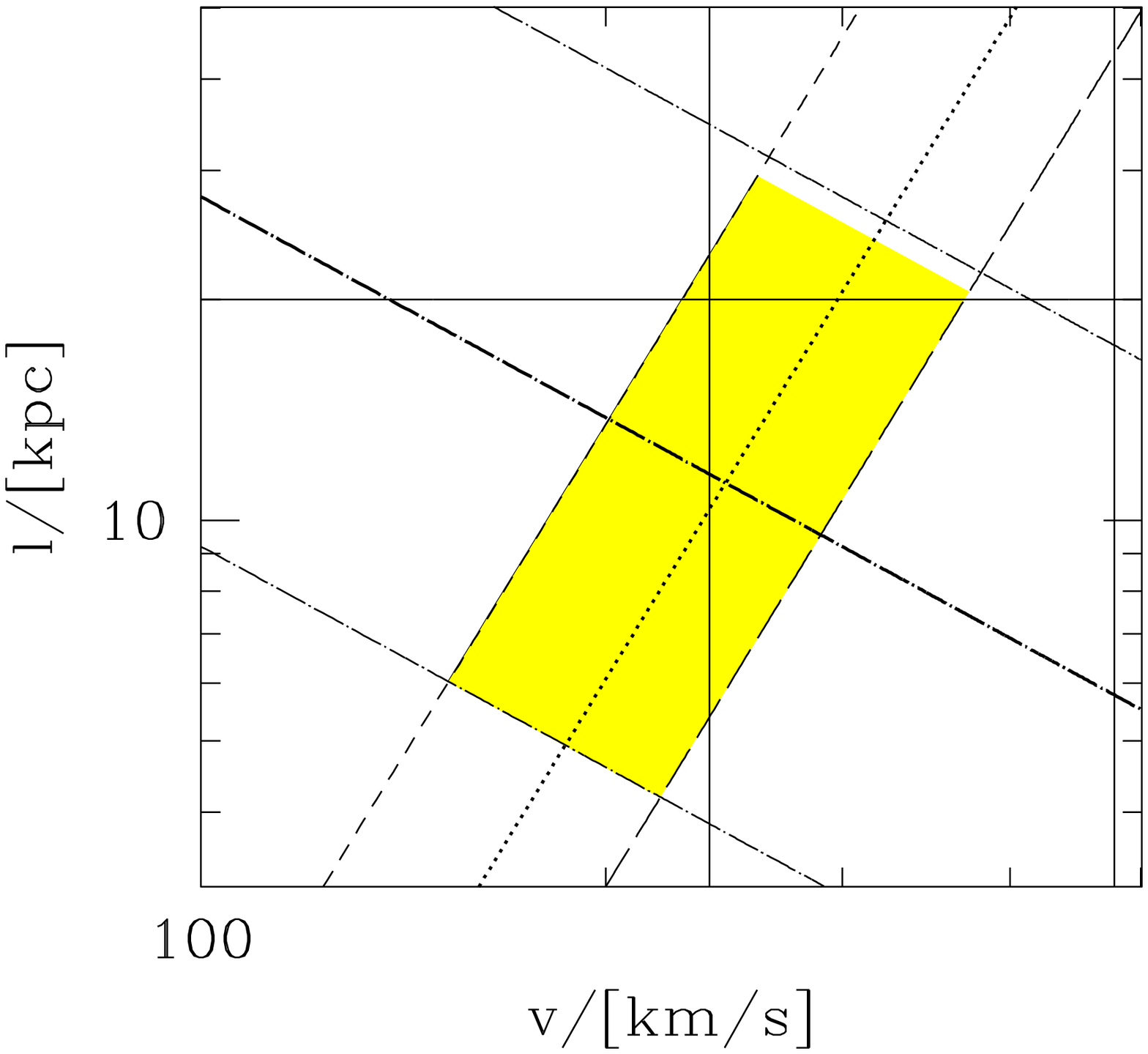}}
   \subfigure[NGC 1550]{
   \label{fig:lv1550}
      \includegraphics[scale=0.27]{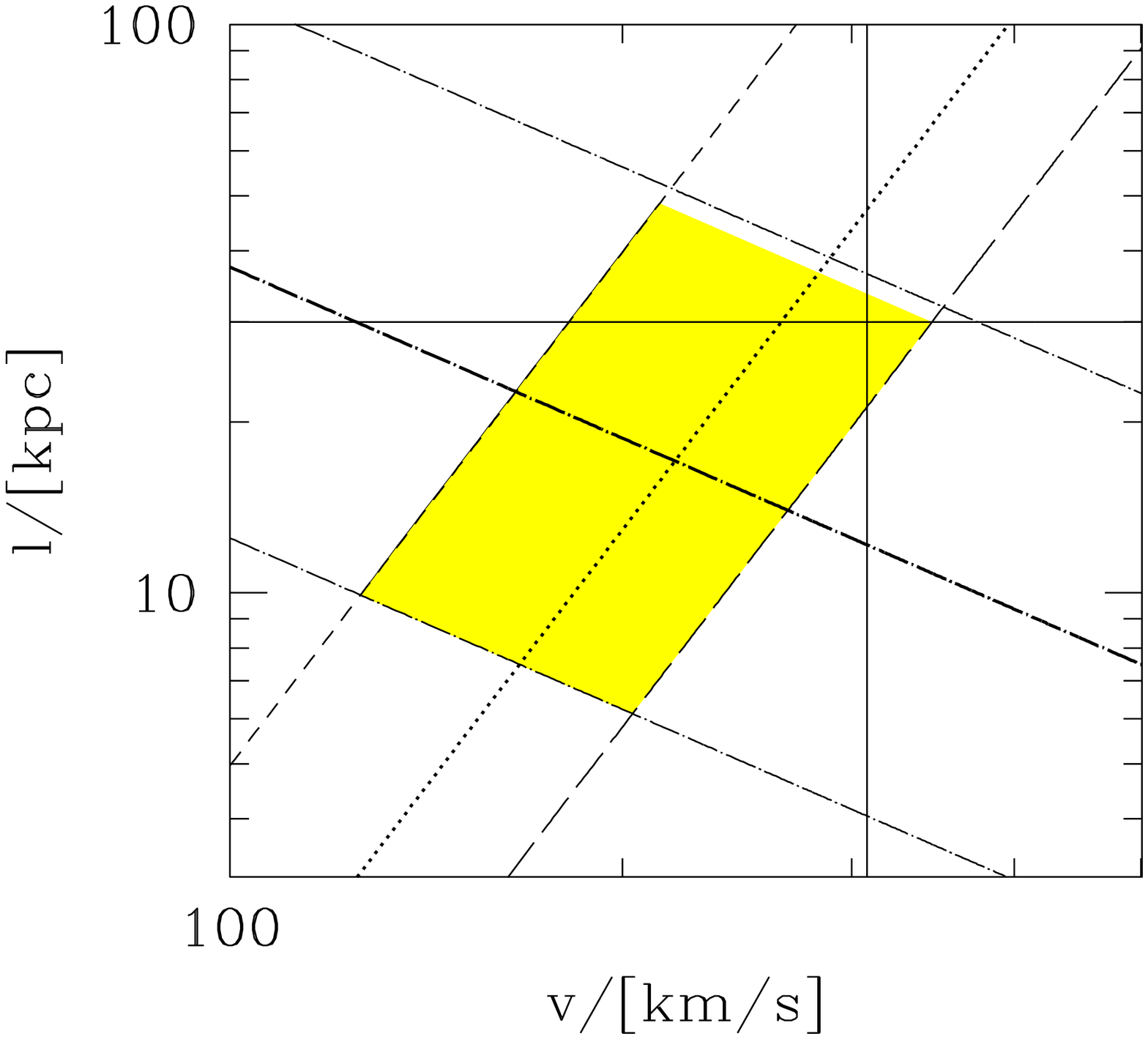}}
    \subfigure[M87]{
    \label{fig:lvM87}
     \includegraphics[scale=0.27]{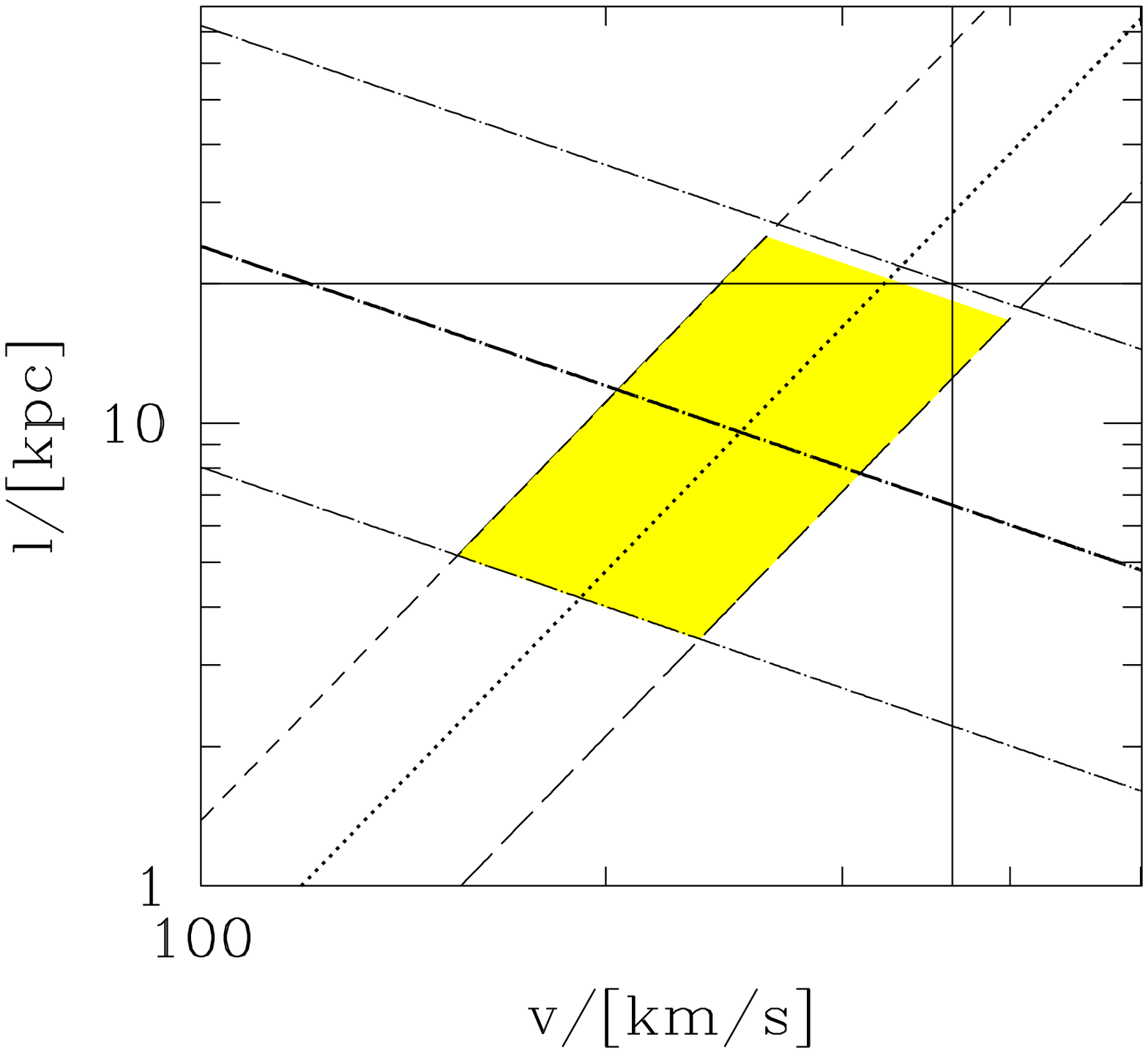}}\\
    \subfigure[AWM4]{
    \label{fig:lvAWM4}
     \includegraphics[scale=0.25]{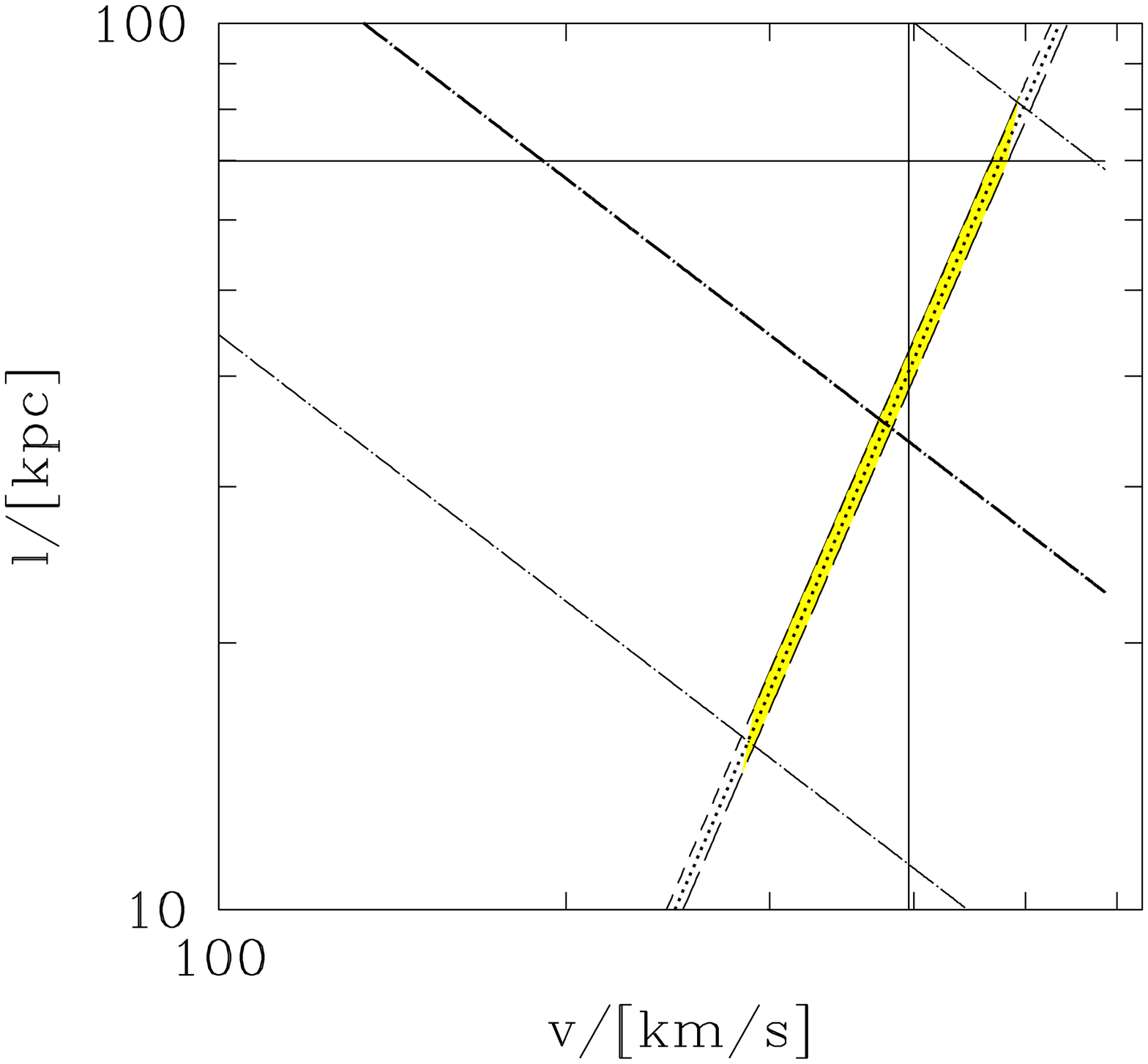}}
    \subfigure[Centaurus]{
    \label{fig:lvcen}
     \includegraphics[scale=0.27]{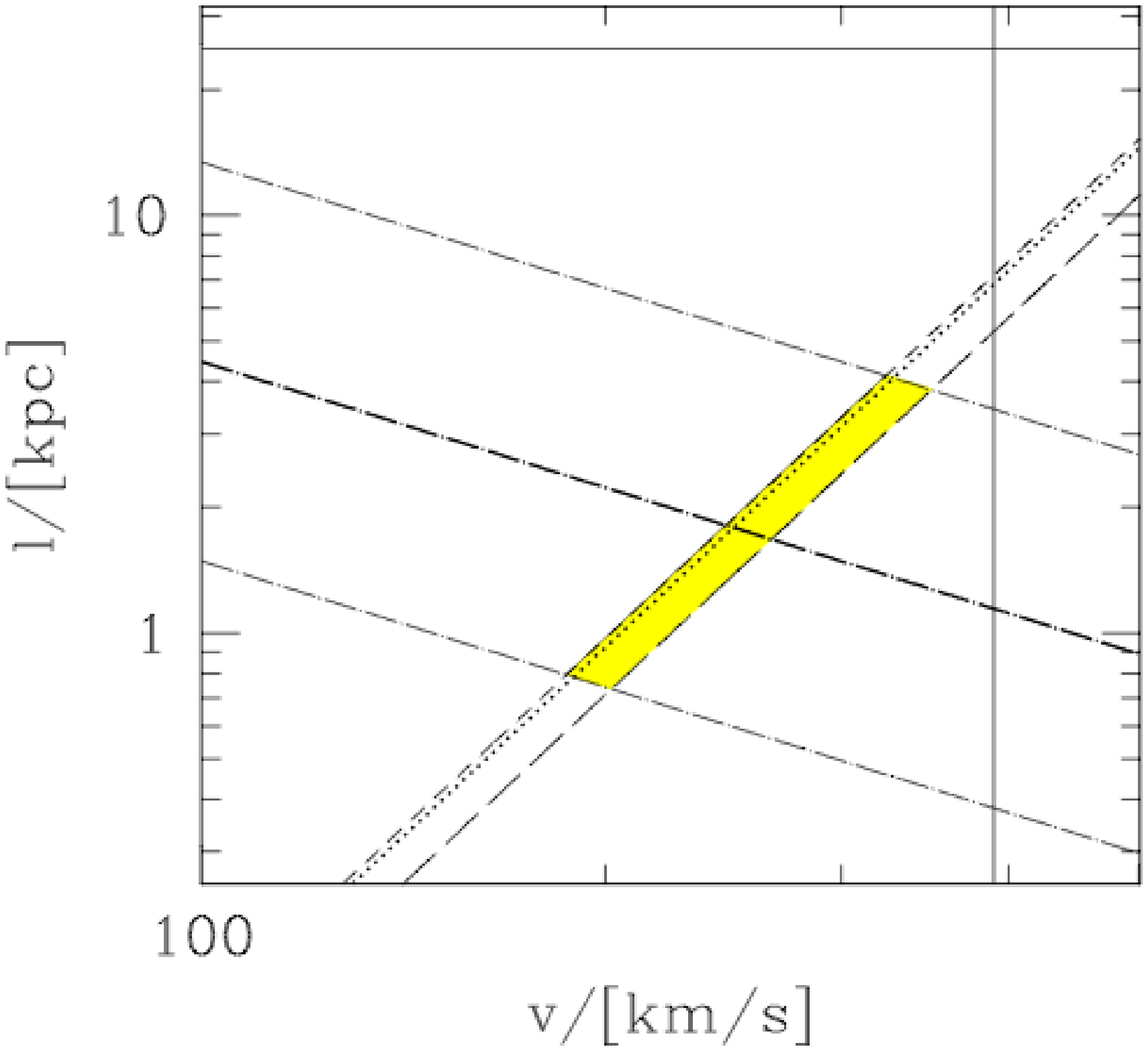}}
    \subfigure[AWM7]{
    \label{fig:lvAWM7}
     \includegraphics[scale=0.25]{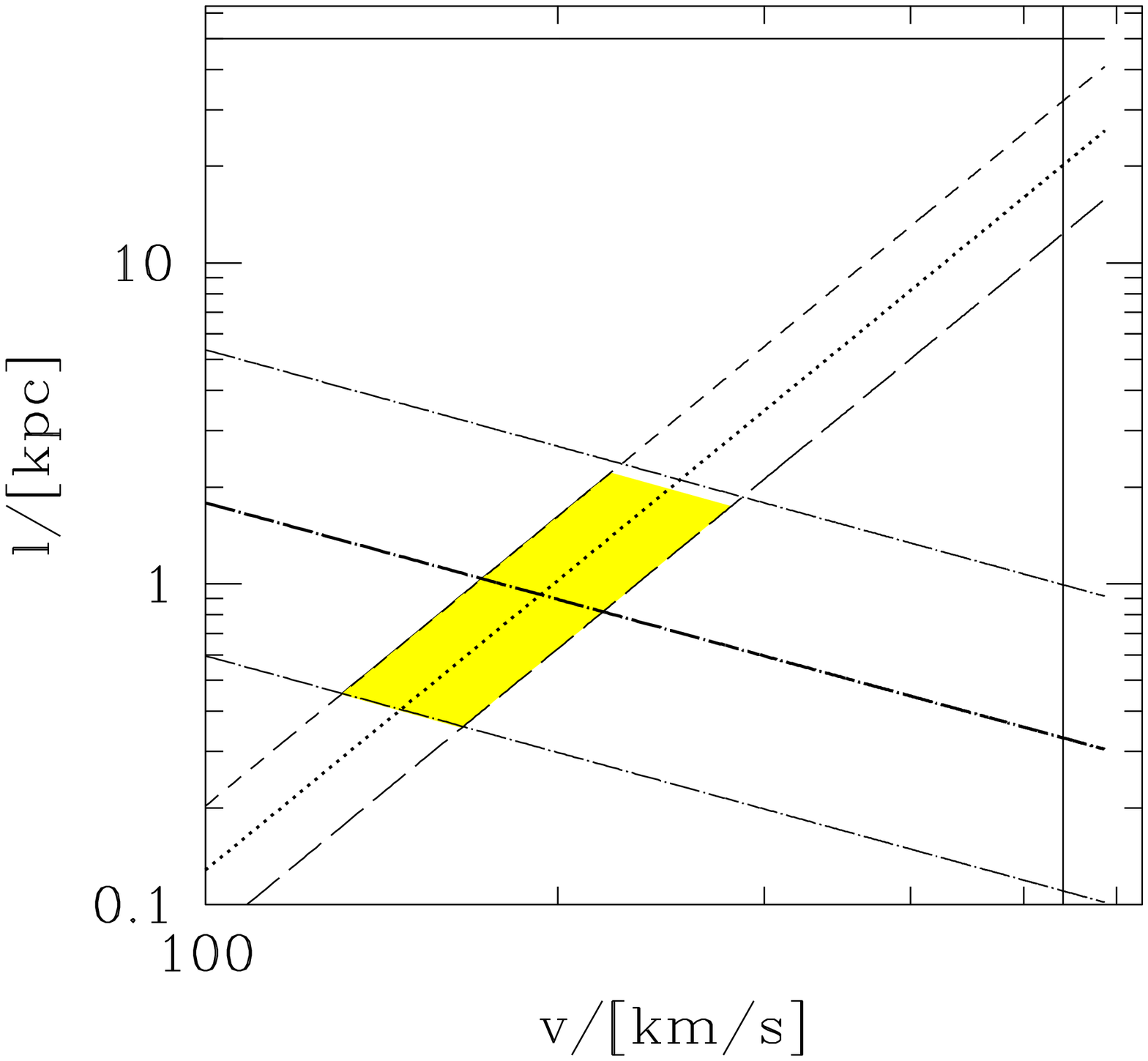}}\\
    \subfigure[A1795]{
    \label{fig:lv1795}
     \includegraphics[scale=0.25]{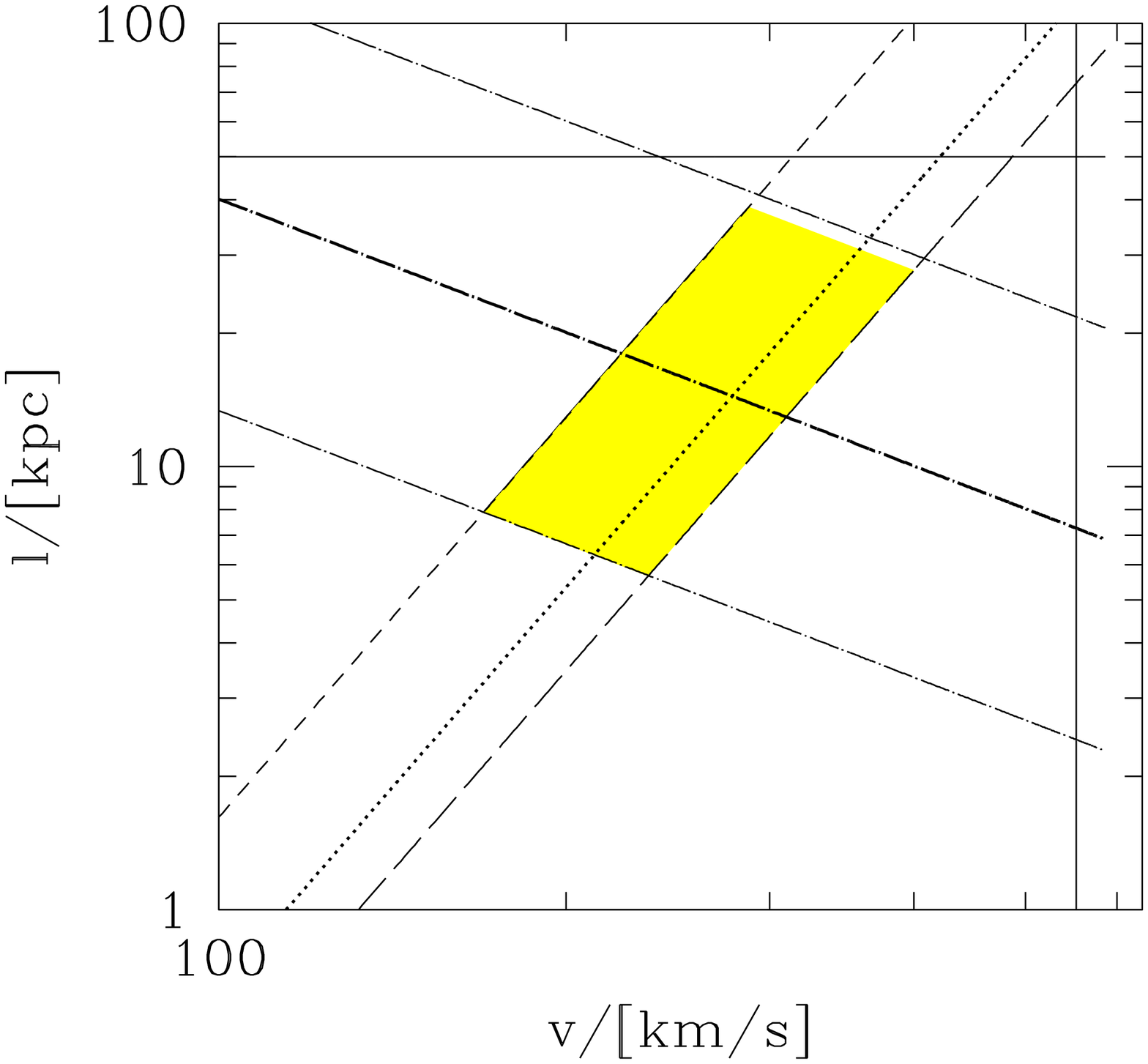}}
    \subfigure[Perseus]{
    \label{fig:lvperseus}
     \includegraphics[scale=0.25]{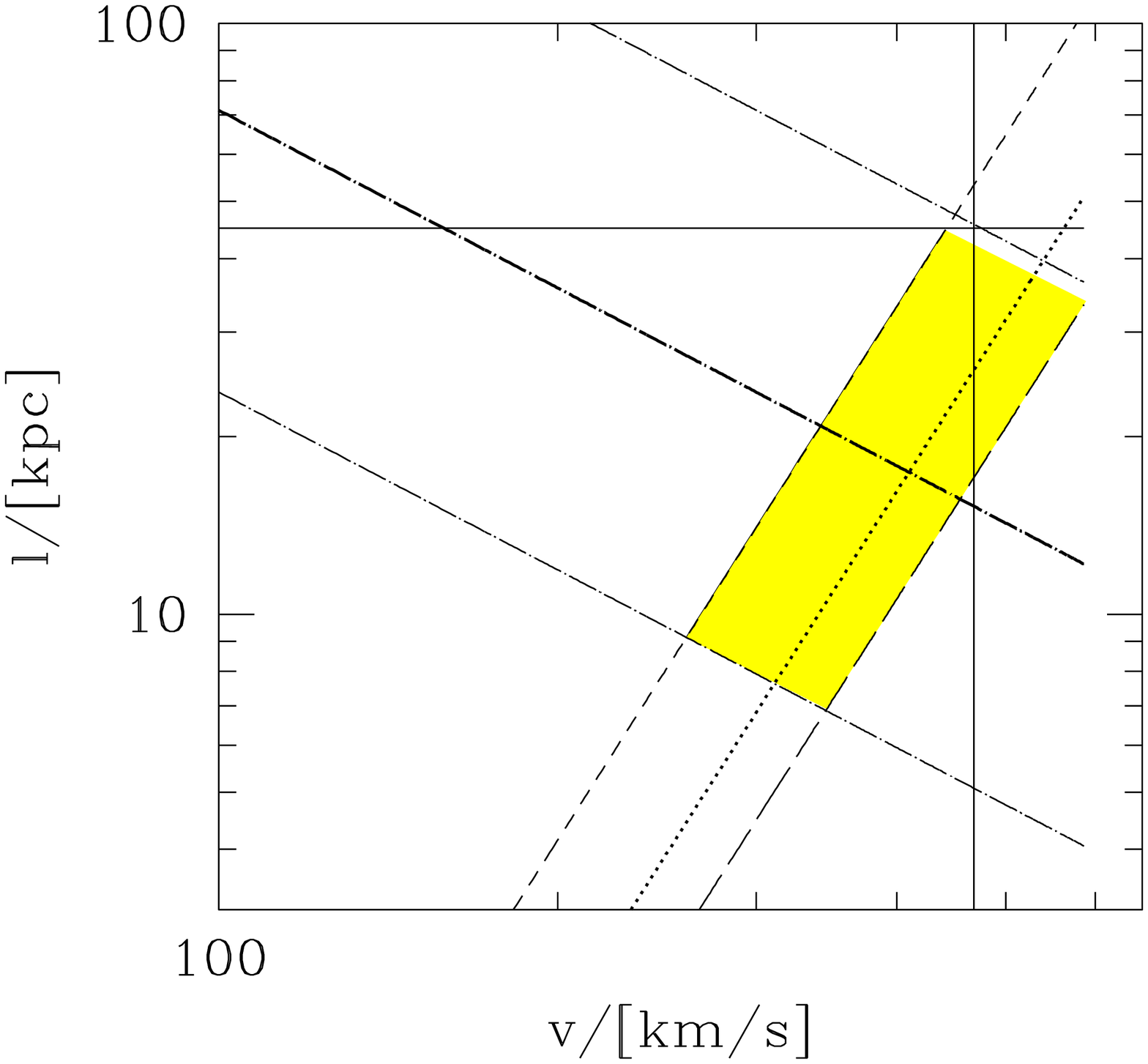}}
\caption{Range of the characteristic velocities $v$ and spatial scales
$l$ of the gas motions which provide the necessary diffusion and
dissipation rates. Along the thick dot-long dashed line the diffusion
coefficient $D=C_1 v l$ is constant and equal to the values listed in
Table $3$, while the thin dot-long dashed lines show the effect of
varying $C_1=0.11$ by factor of 1/3 and 3.  Along the the dotted line
the dissipation rate is equal to the cooling rate at $r_0=r_{cool}/2$;
the dot-short dashed line is for $r_0=r_{cool}$ and the long dashed
for $r_0=r_{cool}/4$ .  At the intersection of these bands, two
conditions are satisfied: i) the gas cooling is balanced by the
dissipation and ii) the diffusion coefficient is of the right order.
The vertical solid lines show the sound speed in the gas $c_s$ for
$r=r_{cool}/2$ and half of it. In the plots where only one line is
present, it corresponds to $c_s/2$.  }
\label{fig:2multifig}
\end{figure}


\twocolumn
\begin{table*}
\centering
\begin{minipage}{177mm}
 \begin{center}
\begin{tabular}{c |c c |c c c| c c c}	
\hline
\hline
Name& $r_{eff}/kpc$ & $L_{TOT}/L_{\odot}^B$           & $r_a/kpc$ & $a(0)/a_{\odot}$ & $b$ & $r_c/kpc$ & $\beta$& $n(0) 10^{-3}/cm^3$\\
\hline
NGC 5044     & $10.05$ [1]   & $4.5 10^{10 }$ [1]     & $58.2$ [2] &$1.07$ & $1.19$    & $15$ [3]  & $0.49$ & $10.66$ \\  
NGC 1550     & $ 9.1$  [4]   & $3.5 10^{10} $ [5]     & $0.8$ [4]& $0.95$ & $0.14$ & $2.5$ [4]& $0.35$ & $7$                \\
$M87^*$        & $7.8$ [5]       & $5.3 10^{10}  $[21]    & $20$ [7][8]& $0.85$ & $0.38$ & $2$ [6] & $0.42$ & $130$ \\
AWM4  &$20$ [9]& $5.5 10^{10}$ [10]& $94$ & $0.78$ [11]& $0.89$ & $76$ [3]& $0.62$ & $3.52$     \\
$Centaurus^*$ & $13$ [9]& $ 5 10^{10}$ [22]& $25$ [13]& $1.31$ & $0.48$ & $7.7 $[14]& $0.57$ & $165$ \\
$AWM7^*$ & $42$ [9] $16.6$ [12]& $ 1.4 10^{11}$ [15]& $7.2$ [16]& $1.6$ & $0.3$ & $3.5$ [17]& $0.25$ & $165$       \\
A1795 &$34$[18]& $2 10^{11}$& $2.7$ [19]& $0.51$ & $0.21$ & $17$ [19][20]& $0.4$ & $50$      \\
Perseus &$15$& $1.6 10^{11}$ &$48$ & $0.54$ & $0.18$ & / & / & / \\     
\hline
\end{tabular}
\end{center}
\caption{Column (1) Name of the object (2)-(3) Effective radius and total  blue luminosity of
 the central galaxy  (4)-(5)-(6) Parameters of the iron abundance $\beta$-profile (7)-(8)-(9)Parameters of the gas density  $\beta$-profile (see the text for the precise form). For the sources marked with ( $*$ ) the best fit was a double $\beta$-model; here we report the parameters of the dominant one. References:[1]Buote et al. 2004[2]Buote et al. 2003[3] Sanderson et al. 2003[4] Sun et al. 2003[5] de Vaucouleurs et al. 1991
 [6] Matsushita et al. 2002[7] Matsushita et al. 2003[8] Gastaldello \& Molendi 2002[9] Schombert 1987[10] Finoguenov et al. 2001[11] O'Sullivan et al. 2005[12]  Bacon et al. 1985
 [13] Fukazawa 1994[14] Ikebe et al. 1999[15] Peletier et al. 1990[16] Furusho et al. 2003[17] Neumann \& B\"ohringer 1995[18] Schombert 1988[19] Tamura et al. 2001[20] Ettori et al. 2002
 [21] Arnaud et al. 1992[22] Laine et al. 2003.}
\end{minipage}
\end{table*}


 This approach of choosing $r_0=r_{cool}/2$ implies that
the cooling rates used will not be drastically different from source to
source (as it would happen if for instance one uses a fixed value of $r_0$
for all the objects in the sample).  
The resulting constraints are shown
in Fig. $2$.  The thick dot-dashed line shows the combinations of $v$
and $l$ which give the same diffusion coefficient, while the thin
dot-dashed lines show the effect of varying $C_1$ by factor of 1/3 and
3.  The effect of varying $C_2$ by factor of 1/3 and 3 was estimated
in Rebusco et al. (2005).  Along the the dotted line the dissipation
rate is equal to the cooling rate at $r_0=r_{cool}/2$; the dot-short
dashed line is for $r_0=r_{cool}$ and the long dashed for
$r_0=r_{cool}/4$ .  The intersection of the two bands gives the locus
of the combinations of $l$ and $v$ such that on one hand the diffusion
coefficient is approximately equal to the required value (Table $3$)
and on the other hand the dissipation rate is approximately equal to
the cooling rate at $r_{cool}/2$.

\subsubsection{Turbulent mixing vs turbulent dissipation}
Since cluster atmospheres are characterized by  positive gradient
of the specific entropy, stochastic motions  should also
lead to the heat flow into the central region (e.g. Dennis \&
Chandran, 2004). Following their work we estimate the rate of heating
due to the turbulent transport of high-specific-entropy
gas into low-specific-entropy regions:
\begin{equation}
\Gamma_{tt}=\nabla \cdot (D \rho T_{gas} \nabla e),
\label{eq:gtt}
\end{equation}
where $T_{gas}$ is the temperature and $e=C_V
ln(c_s^2/\gamma\rho^{\gamma-1})$ is the specific entropy (adiabatic
index $\gamma=5/3$, specific heat at constant volume $C_V=3k/2\:\:
n_H/\rho$ ).

\begin{figure}
\plotone{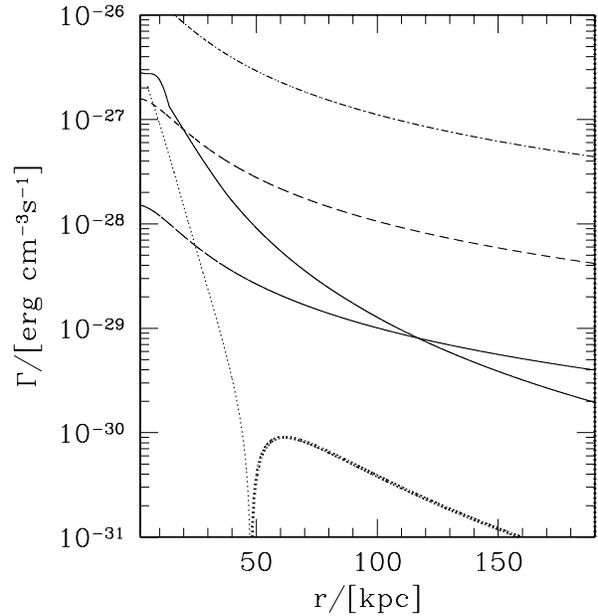}
\caption{NGC 5044: the solid line shows the cooling rate as a function of 
radius; the dotted line is the heating rate $\Gamma_{tt}$ evaluated
for $D=9\times10^{28}$cm$^2$~s$^{-1}$: the thin line corresponds to
the part where $\Gamma_{tt}$ is positive, while the thick one is for
$\Gamma_{tt}<0$. The remaining lines represent the dissipation heating
rate $\Gamma_{diss}$ for the same $D$, but different velocity and
length scales: $l=l_{ref}=11$ kpc and $v=v_{ref}=245$ km/s
(short-dashed), $l=l_{ref}/1.8$ and $v=1.8\:v_{ref}$ (dot-dashed),
$l=1.8\:l_{ref}$ and $v=v_{ref}/1.8$(dot-dashed).  For the other sources,
these plots show a similar qualitative behavior. }
\label{fig:rate}
\end{figure}

First we consider the case of a diffusion coefficient independent of
the radius.  A typical example  of $\Gamma_{tt}$ radial
dependence (NGC 5044), for the diffusion coefficients taken from Table 3, is shown
in Fig. \ref{fig:rate}. While the heat flow is always towards the
center, the volume heating  rate $\Gamma_{tt}$ is positive inside a
certain radius and it becomes negative (i.e. cooling) in the outer
regions. This is of course an expected result - for constant
diffusion coefficient the heat flux $r^2\:D\: \rho\: T_{gas}
\drv{e}{r}\sim r\:\rho$ (see eq.\ref{eq:gtt}) and there is likely to
be a radius near which this function is almost independent from
the radius.  As it is seen in Fig. \ref{fig:rate} the turbulent transport (dotted
line) calculated for $D=9\times10^{28}$ cm$^2$~s$^{-1}$ is an important
source of heat in the innermost 10 kpc, but $\Gamma{tt}$ quickly declines with
the radius. Obviously the diffusion coefficient has to increase with
the radius if the turbulent heat transport $\Gamma_{tt}$ balances the cooling
over a broad range of radii. For comparison the heating rate due to
the dissipation of turbulent motions $\Gamma_{diss}$ (for the
parameters taken from Table 3) is shown by the dashed line in
Fig. \ref{fig:rate}. Note that the rate of dissipation is a
very strong function of $v$ and  variations in $v$ by a
factor of 2 (for a fixed diffusion coefficient $\propto l\times 
v$) cause  variations of $\Gamma_{diss}$ by a factor of 16 (see
Fig. \ref{fig:rate}). 

Chandran (2005) considered convective motions in the cluster cores
driven by the cosmic rays injected by a central AGN. In his two-fluid
(thermal-plasma and cosmic-ray) mixing-length theory approach, the
characteristic length and velocity scales are functions of the 
radius. Assuming that the turbulent heat transport is the dominant
source of heat we calculated the diffusion coefficient $D_0(r)$
(needed to balance heating and cooling at every radius) by integrating
the equation $\Gamma_{tt}=\Gamma_{cool}$ over the radius.  The resulting
diffusion coefficient linearly increases with the radius and it reaches the
value of $D=9\times10^{28}$cm$^2$~s$^{-1}$ at $r\sim$ 5 kpc. Such
radially dependent diffusion coefficient produces progressively more
and more efficient mixing with increasing distance and it leads to an
abundance profile which drops to $\sim$0.1 already at $\sim$10 kpc
(see Fig. 4).

We next assumed that $\Gamma_{diss}\approx \Gamma_{cool}$ at every
radius and we calculated the quantity $v^3/l$ as a function of the radius. We
then considered several possibilities for the radial dependence of $v$
and $l$ and we derived the corresponding diffusion coefficients.

\begin{eqnarray}
D_1(r)=&c_1 v l(r), &v={\rm const} \\
D_2(r)=&c_1 v(r) l, &l={\rm const} \nonumber\\
D_3(r)=&c_1 v(r) l(r), &l(r)=0.3 r.\nonumber 
\end{eqnarray}

\begin{figure*}
\plottwo{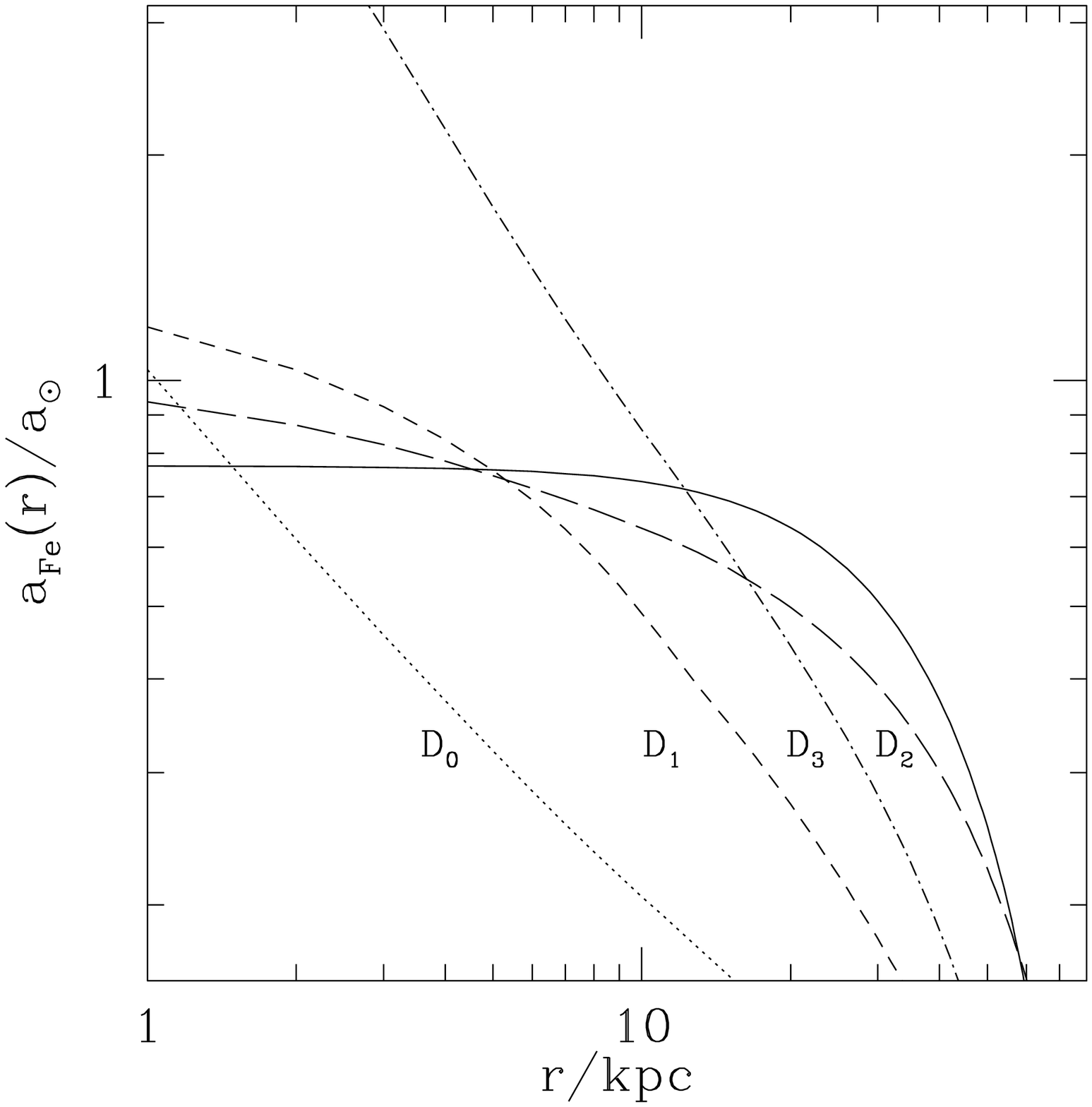}{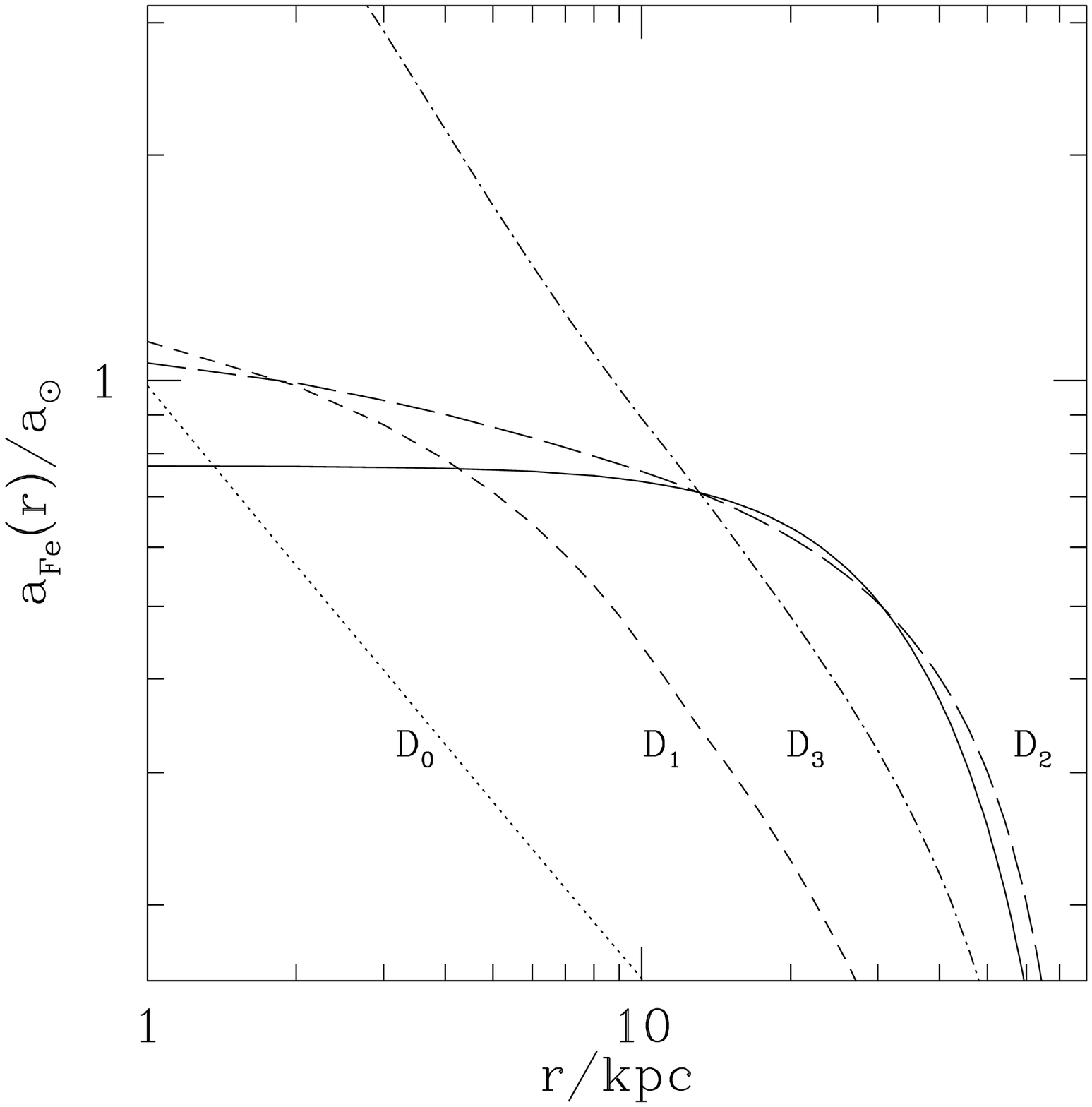} 
\caption{NGC 5044: in both plots the solid line represents the observed abundance profile from which
 the constant value $a_b=0.3 a_\odot$ is subtracted. The dotted line is for $D=D_0(r) $, the short-dashed
 for $D=D_1(r) $, the long-dashed for $D=D_2(r) $ and the dotted-dashed for $D=D_3(r)$(see subsection $3.4.2$ for an explanation):
 they show the expected abundances obtained according to equation $1$, starting  from a flat
 zero profile (left panel, $\tau_{age}=5$ Gyr) and starting from the present-day profile (right panel, 
after $4$ Gyr from now). For the other sources, these plots
 display a similar qualitative behavior.}
\label{fig:3multifig}
\end{figure*}

The choice of $l(r)\propto r$ for $D_3(r)$ is motivated by the work of
Chandran (2005). For the first two models the constant $v$ and $l$
were fixed at the values taken from Table 3. The abundance profiles
obtained by integrating equation $1$ using these diffusion coefficients
are shown in Fig. \ref{fig:3multifig}. The case of $D_2(r)$
(i.e. constant length scale $l$ and variable velocity scale $v$) fits
the observed profile the best, while the other versions of $D(r)$ yield
profiles which strongly deviate from the observed one.

From these plots one can conclude that it is possible to construct a
model with radially dependent velocity and length scales such that
the abundance profiles are approximately reproduced and that the dissipation of
turbulent motions compensates the gas cooling over a wide range of
radii. In the simplest model of this type ($D_2(r)$ above) the length
scale is independent of the radius and the velocity scale is a slowly
decreasing function of the radius.


\section{Results and their discussion}
\label{sec_results}
In Table $3$ the estimated values of the diffusion coefficient $D$,
the characteristic velocity $v$ and the spatial length scale $l$ are
listed. The set of the parameters for the iron enrichment models
(listed in Table $4$, described in the Subsection $3.3$) are chosen in
order to get a reasonable approximation of the observed profiles (see
Fig. 1): a factor of $3$ smaller (or larger) diffusion coefficients
produce too peaked (or too shallow) abundance profiles, inconsistent
with the data.

For each source we made a consistency check evaluating the
characteristic value of the diffusion coefficient by comparing the
half mass radius of the observed central excess and the excesses
produced by ejection and diffusion in the model. For small diffusion
coefficients the metal distribution should follow the light
distribution of the central galaxy (i.e. have the same effective
radius), while for the increasing diffusion coefficient the effective
radius should also increase. The values derived this way were
consistent with those given in Table 3.

We then used the observed abundance peaks in each object as initial
conditions and we calculated the subsequent evolution of the abundance
profiles by  fixing $D$ at the value from Table 3. In each
case 3 times smaller (or larger) diffusion coefficient causes a quick
steepening (or flattening) of the abundance peaks on time scales of a
Gyr, while for diffusion coefficients of the order of those listed in
Table $3$ the evolution is very slow on these time scales.


\begin{table}
\begin{center}
\begin{tabular}{c |c c c c c}
\hline
\hline
Name	    & $r_{cool}/2$ kpc &$D$ cm$^2$~s$^{-1}$          & $l$   kpc   & $v$ km~s$^{-1}$ & $v/c_s$ \\
\hline
NGC 5044     & $20$ &$9\times10^{28}$   &$11$	&$245$ & $0.51$    \\
NGC 1550      & $30$ & $1.2\times 10^{29}$ &$17$   & $219$ & $0.35$            \\	
M87           & $20$ &$8\times 10^{28}$  &$10$   &$250$ & $0.34$           \\
AWM4          & $70$ &$4.5 \times 10^{29}$ &$35$   &$378$ & $0.45$           \\
Centaurus   & $25$ & $2 \times  10^{28}$ & $2$   &$250$ & $0.30$    \\ 
AWM7           & $50$ &$6  \times 10^{27}$ &$1$    &$193$  & $0.18$   \\
A1795          & $50$ & $1.3\times 10^{29}$     &$14$   &$278$ & $0.25$ \\
Perseus       & $45$ & $2.4\times 10^{29}$   &$17$   &$410$ & $0.44$\\
\hline
\end{tabular}
\end{center}
\caption{Estimate of the turbulence parameters for each source in our sample: name (1), half the cooling
 radius (2), diffusion
 coefficient (3), length scale (4) and velocity (5-6) of the gas motion.}
\label{tab_results}
\end{table}


The diffusion coefficients listed in Table 3 are mostly of the order
of $10^{29}$ cm$^2$~s$^{-1}$, apart from those obtained for AWM4,
Centaurus and AWM7 (see next subsections). These values can be
considered as an an upper limit on the effective diffusion coefficient
set by the random gas motions, since other processes may contribute to
spreading the metals through the ICM. 

Note that our simple model predicts the abundance to be the highest in
the center of the cluster. Thus the abundance "hole" observed in
several well studied clusters (e.g. B{\"o}hringer et al. 2001, Schmidt
et al. 2002) contradicts this conclusion. Potentially there are
several explanations of the abundance hole. The simplest one supposes
that the spectral models used for the abundance determination are not
adequate for the central region, where a complicated mixture of
different phases is present (e.g. Buote 2000). The resonant scattering
and the diffusion of the line photons to larger radii is another
possible explanation (e.g. Gilfanov et al. 1987, Molendi et
al. 1998). However the model advocated here assumes that the gas in
the center is not at rest, but it is involved in complex pattern
of motions, which should reduce the effective optical depth
(e.g. Churazov et al. 2004). Velocities of order the of few hundred
km/s should make most of the lines optically thin. Less
straightforward interpretations of the abundance hole are also
possible, but they are beyond the scope of this paper. Future high
energy resolution observations with micro-calorimeters should be able to
resolve this question.

The sizes of the turbulent eddies obtained in this work are of the
order of $10$ kpc and the characteristic velocities of the order of
few 100 km/s.  In our model the velocity scale is determined from the
combination of two quantities $D\sim vl$ and $\Gamma_{diss}\sim
v^3/l$: $v \sim (D\: \Gamma_{diss})^{(1/4)}$. As a result the
estimates of $v$ come out very similar for all the objects, while $l$
spans over a larger interval, compensating for all the peculiarities.
We stress that these estimates have to be taken with caution, due to
the simplicity of the model and the to uncertainties in the enrichment
process.

\subsection{AWM4}
\label{subsec:4}
This object is not a cooling flow cluster, unlike the other objects in the
sample. It does contain however a central bright galaxy (NGC 6051)
which should enrich the ICM with metals in the same way as in cooling
flow clusters. The enrichment will be even more efficient in AWM4
because the gas density is relatively small and even a modest amount
of metals should produce a prominent abundance peak. As one can see
from Table 3 the diffusion coefficient required to explain the lack of
such abundance peak in AWM4 is larger than for any other objects in
the sample. The length scale is also the largest in the sample. This
can be easily understood since in this source a large value of the
diffusion coefficient is needed along with a modest heating rate. As
it is clear from eq.(5) and (6) this can be most naturally done by
setting a large value of the gas motions spatial scale $l$.

If we would like to stay within the frames of our model then this is
the object with the strongest level of  gas mixing. There is a
powerful radio source in NGC 6051 which clearly interacts with the ICM
(e.g. O'Sullivan et al. 2005). It is not obvious however if the
central AGN alone is responsible for the gas mixing and for the
appearance of AWM4 as a non-cooling flow cluster. The alternative
would be a recent merger followed by an intensive mixing of the ICM,
but this case seems to be excluded by observations (O'Sullivan et
al. 2005).

\subsection{Centaurus}
\label{subsec:cen}
The Centaurus cluster has a central abundance peak which is difficult
to reproduce with the enrichment model presented above. In order to
produce such a large amount of iron in the central excess of this cluster, a
long accumulation time (of the order of $10$ Gyr) and a high SNIa
explosion rate are necessary (see Table 3). With these parameters we
get a diffusion coefficient of $\sim 1.5\times 10^{28}$
cm$^2$~s$^{-1}$.

A previous estimate by Fabian et al. (2005) obtains that the diffusion
coefficient in Centaurus is lower than $6\times10^{28}$
cm$^2$~s$^{-1}$, consistently with our findings. Our result is also
consistent (within the uncertainties of the profiles) with the value
found by Graham et al. (2005).  Since Graham et al. (2005) use higher
blue luminosity for the central galaxy, their choice of enrichment
parameters is less extreme than our.  They also adopted higher value
of the central abundance peak ($a_0\sim 2 a_\odot$), derived from
fitting the spectra with a cooling flow model, while our lower central
abundance ($a_0\sim 1.3 a_\odot$) is based on the two-temperature fits
by Sanders \& Fabian (2002).  For their choice of parameters supersonic
velocities are required to satisfy the equations (5) and (6).  For
lower luminosity (see Table 2), subsonic solutions are possible.
However the results for Centaurus cluster remain quite controversial.

One plausible explanation of the extremely steep abundance gradient in
Centaurus is the recent disturbance of the gas in the cluster
core. "Cold fronts" are observed in many clusters with cool cores
(e.g. Markevitch et al. 2000).  These features are believed to form
when the gas sloshing in the potential well brings layers of gas with
different entropies in close contact: in a similar fashion layers of
gas with different abundances can form a transient feature
characterized by a steep abundance gradient. In the frame of the model
discussed here any steep abundance gradient translates into strong
limits on the diffusion coefficient.


\subsection{AWM7}
\label{subsec:7}
Taking the results of our analysis at face value, this cluster does
not require strong stochastic motions to match the observed and
predicted abundance profiles. There is also no evidence for a strong
radio source in it(e.g. Furusho et al. 2003). From this
point of view it is tempting to conclude that AWM7 is now passively
cooling and that the activity of the central AGN is only about to
start (in another cooling time $\sim 7\times10^8$ years). Then our
estimate of the velocity and the length scale based on the assumption
of a balance of the gas cooling would overestimate the combination
$v^3/l$ and underestimate $l$.  Note however that Furusho et
al. (2003) reported an unusual substructure in the metal distribution in
the AWM7 core, which is at odds with the assumption that gas motions
are almost absent in this cluster. Given that in several clusters an
abundance hole is observed it is possible that the substructure in
AWM7 has a similar nature.

\subsection{Radio bubbles and stochastic gas motions}
Vazza et al. (2006) simulated non-cooling flow clusters, fin\\
ding that
the energy due to turbulent bulk motions scales with the mass of the
cluster: similar calculations had been done by Bryan \& Norman (1998).
In these simulations the turbulence is a result of the cluster
formation, while in the core of CF clusters one can expect that the
AGN is playing a dominant role in driving the turbulence: as a
consequence the mass scaling of the two processes may be different.

Allen et al. (2006) found recently a tight correlation between the
estimates of the Bondi accretion rate on to the central black hole and
the jet power (estimated from the observed expanding bubbles) in X-ray
luminous elliptical galaxies (similar calculations have been
previously done by Di Matteo et al. 2001, Churazov et al. 2002 for M87
and by Taylor et al. 2006 for NGC 4696). At the same time the
comparison of the jet powers and the cooling rates (e.g. B{\^i}rzan et
al. 2005) also shows a reasonable agreement. These results suggest
that the amount of energy provided by jets (in the from of bubbles of
relativistic plasma) is regulated by the gas properties in such a way
that the gas cooling is approximately compensated. One can compare the
rough estimates of the turbulence length scales (of the order of few - few tens kpc
, see Table 3) with the observed sizes of radio bubbles in cluster cores
(e.g. Dunn \& Fabian 2004, Dunn, Fabian \& Taylor 2005, Churazov et
al. 2001). Broadly the numbers are of the same order.  E.g. the
observed radio bubbles sizes in M87 vary from $\sim 1$ to $\sim 10$
kpc (depending if one refers to the inner lobes or to the torus-like
eastern or the outer southern cavities). In the Perseus, A1795 and
Centaurus clusters the observed bubbles sizes also fall in the range
few - few tens kpc. The characteristic velocities of the bubbles
motions are of the order of $\sim 0.5 c_s$ and they are also broadly consistent
with the values listed in Table 3. From this point of view the
observed bubbles seem to have all the necessary characteristics required
to drive the gas motions needed to spread the metals.

\section{Conclusions}
We show that the iron abundance profiles observed in a small sample of
nearby clusters and groups with cool core are consistent with the
assumption that metals are spreading through the ICM with 
effective diffusion coefficients of the order of
$10^{29}$cm$^2$~s$^{-1}$. This value does not show any obvious trend
with the mass or the ICM temperature of the system.

For characteristic velocities of the order of $300$ km/s and length
scales of the order of $10$ kpc, the dissipation of the gas motions
would approximately balance the gas cooling. When observations are
available, our estimates of the turbulent length scales are consistent
with the observed buoyant bubbles size, pointing at the central
supermassive black hole as the likely origin of the core gas motions.

Two objects in our sample require very different diffusion
coefficients: for AWM7 very small level of gas mixing is needed, while
for AWM4 the mixing has to be very intense. It is possible
that these two objects hint to the intermittence of the gas mixing and that
they represent two different episodes of the cluster evolution: AWM7
is only weakly disturbed now (but it will likely be disturbed soon by the onset
of AGN activity), while AWM4 has been already mixed recently.
O'Sullivan et al. 2005, using the analogy with M87 (where an abundance
gradient is present in spite of clear signs of a radio jets interaction
with the ICM) suggested that it is unlikely that AGN activity is
responsible for the absence of the abundance peak in AWM4. One cannot
however completely exclude the possibility that in the past there was a period 
of truly violent AGN activity in AWM4, which not only removed the
central abundance peak, but also destroyed the whole cooling flow.

\begin{table*}
\centering
\begin{minipage}{177mm}
 \begin{center}
\begin{tabular}{c |c c c |c c c|c}	
\hline
\hline
Name & $a_b/a_\odot$ & $M_{Fe}/M_\odot \times 10^8$ & $r_m/kpc$ & $k$
& $SR/SNU$ & $t_{age}/Gyr$ & $D/[cm^2/s]\times 10^{28}$\\
\hline
NGC 5044   &  $0.2$ & $1.5$ & $42$ & $1.1$ & $0.26$ & $8$ &$12.2^*$  \\
                   & \# $0.3$ & $0.5$ & $36$ & $1.1$ & $0.15$ & $5$ & $8.7$ \\
                   & $0.4$ & $0.8$ & $31$ & $2$    & $0.06$ & $8$ & $9$ \\
NGC 1550   & \# $0.3$ & $0.8$ & $67$ & $1.1$ & $0.15$ & $7.6$ & $12.6^*$ \\ 
M87            & \# $0.2$ & $0.9$ & $46$ & $1.1$ & $0.15$ & $6.2$ & $8.3$ \\
                   & $0.3$ & $0.8$ & $29$ & $1.1$ & $0.24$ & $5$  & $8.8$ \\
AWM4         & \# $0.2$ & $7.2$ & $83$ & $ 2$   & $0.33$ & $10$ & $45$\\
                   &  $0.3$ & $3.8$ & $64$ & $1.1$ & $0.34$ & $10$ & $37.5$ \\
                   & $0.4$ & $1.9$ & $55$ & $2$    & $0.30$ & $6.4$ & $25^*$\\ 
Centaurus   & \# $0.3$ & $6.0$ & $47$ & $2$    & $0.33$ & $10$ & $2^*$ \\
                   & $0.4$ & $3.6$ & $37$ & $1.1$ & $0.39$ & $10$ & $1^*$\\ 
AWM7         & \# $0.3$ & $7.6$ & $50$ & $1.7$ & $0.15$ & $10$ & $0.6^*$ \\
                   & $0.4$ & $3.4$ & $32$ & $1.1$ & $0.15$ & $8.7$ & $0.5$ \\ 
A1795         & $0.2$ & $16$ & $140$ & $2$   & $0.30$ & $8.4$ & $36$\\
                   & \# $0.3$ & $2.9$ & $62$ & $1.1$ & $0.15$ & $5.4$ & $12.5$ \\
\hline
\end{tabular}
\end{center}
\caption{Enrichment models:column (1) Name of the object (2) Total iron excess; the hash 
indicates the model that was used to produce the plots in Fig. 1 and 2 (3) Half mass radius (4) $k$ (5) Present-day
 SNIa rate (6) Source lifetime (For the definitions see the text)-(7)
Effective diffusion coefficient: the models marked with the asterisk have an
uncertainty due to the choice of the criterion to fix
 $D$ of a factor $\%/\times 2$, while for the other models the
		  uncertainty is
 smaller.}
\end{minipage}
\end{table*}

\section*{Acknowledgments}
P.R. is grateful to the  International Max Planck Research School for its support
 and to R.Mushotsky and his group at GSFC for their hospitality.
We thank an anonymous referee for useful suggestions.

\label{lastpage}
\end{document}